\documentclass{article}

\usepackage[left=40mm, right=40mm, top=40mm, bottom=30mm]{geometry}

\usepackage[utf8]{inputenc}
\usepackage{amsmath,amssymb}
\usepackage[english]{babel}
\usepackage{blindtext}
\usepackage{amsthm}
\usepackage{comment}
\usepackage[titletoc,title]{appendix}

\usepackage{url}
\RequirePackage{color,graphicx}
\usepackage{hyperref}
\definecolor{linkcolour}{rgb}{0,0.2,0.6}
\hypersetup{colorlinks,breaklinks,urlcolor=linkcolour, linkcolor=linkcolour}

\usepackage{thmtools}
\usepackage{thm-restate}
\usepackage{cleveref}
\newtheorem{theorem}{Theorem}%[section]
\newtheorem{lemma}[theorem]{Lemma}
\newtheorem{corollary}{Corollary}[theorem]
\theoremstyle{remark}

\theoremstyle{definition}
\newtheorem{definition}{Definition}%[section]

\DeclareMathAlphabet{\mathpzc}{OT1}{pzc}{m}{it}

\title{
General Mathematical Proof of Occam’s Razor;
\\
Upgrading Theoretical Physicists' Methodology
}
\author{%
Gabriel Leuenberger\thanks{%
For correspondence: \href{mailto:g@leuenberger.ai}{g@leuenberger.ai}%
}%
}
\date{2025}

\begin{document}

\maketitle

\begin{abstract}
This paper's first aim is to prove a modernized Occam's razor beyond a reasonable doubt.
To summarize the main argument in one sentence:
If we consider all possible, intelligible, scientific models of ever‐higher complexity, democratically, 
the predictions most favored by these complex models will agree with the predictions of the simplest models.
This fact can be proven mathematically, thereby validating Occam's razor.

Major parts of this line of reasoning have long preexisted within the depths of the algorithmic information theory literature, 
but they have always left room for doubts of various kinds.
Therefore, we increase the generality, completeness, clarity, accessibility, and credibility of these arguments by countering over a dozen objections.
We build our mathematical proof of Occam's razor on the shoulders of the exact ‘chain rule’ for Kolmogorov complexity.

Concerning physics, we then go on to diagnose the primary amendable root cause of the present stagnation of the research field of fundamental theoretical physics.
We show that the effective antidote would consist in a practically feasible upgrade to the theoretical physicists' research methodology:
When proposing new theoretical models, 
physicists should simply calculate and report the total amount of information that their models consist of.
We explain why this methodology would be highly effective as well as how these calculations could be performed efficiently. 
\end{abstract}

%\keywords{metascience, crisis in physics, metamathematical regularization}

\section{Introduction}
Why exactly should Occam’s razor be true and why should this be relevant to the stagnant field of fundamental theoretical physics?
In this paper, 
we provide a mathematical understanding, accessible to scientists of any field,
on how and why exactly Occam's razor is true, 
i.e., why it greatly increases scientists' chances of arriving at correct novel predictions. 
Based on these insights, 
we then go on to explain how one ought to combat the infamous stagnation of the research field of fundamental theoretical physics in a highly effective and practically feasible manner,
via the introduction of a new methodological quality standard for writing papers, 
which would consist in the calculation and the reporting of exactly how complex any novel scientific model is.
We explain this practical methodology in \mbox{Subsection \ref{alleviate}}.
But to return to the more abstract, 
general mathematical proof of Occam's razor,
the following paragraph is a short summary of the proof's main line of reasoning: 

In science, models are employed to explain past observations and to predict future observations,
e.g., future experimental outcomes.
For all‐encompassing generality, we are concerned with all fully intelligible models, 
i.e., all models that one could express in a Turing-complete language.
Any past observation can be explained by different models, by infinitely many models, in fact.
This can lead to an infinite number of different predictions based on the very same past observation, making it seemingly impossible to select the correct prediction.
Note however, that not all models will contradict each other on their predictions.
All practical feasibility aside, one reasonable method, therefore, would be to select the correct prediction via a ‘fair democracy’ of models, 
where we elect the prediction supported by the largest number of models.
Note that the overwhelming majority of such ‘voters’ will be highly complex models, as more of them exist for obvious combinatorial reasons.
Remarkably, however, despite all this charitability toward complex models, 
it can be proven mathematically that the outcome of such an election will nevertheless agree with the simplest models,
which goes to prove Occam's razor.

This short summary may already have been raising various questions that are to be answered in subsequent sections.
In \mbox{Section \ref{history}}, 
we provide a brief overview of the history of Occam's razor.
In \mbox{Section \ref{model complexity}}, 
we establish an objective foundation for a general measure of the complexity of scientific models.
In \mbox{Section \ref{main section}}, 
we present the main argument for Occam's razor, 
including our proofs of the supporting mathematical theorems,
and we gather and discuss any initial assumptions that we have introduced throughout our paper.
In \mbox{Section \ref{Physics section}},
we explain various issues related to the field of fundamental theoretical physics, such as its stagnation and how to upgrade to the methodology of the theoretical physicists.
Finally, in \mbox{Section \ref{conclusion}}, 
we conclude with our key takeaways and our future outlook.
The twelve main objections that we address throughout our paper are marked by Roman numerals.

\section{History of Occam’s Razor}\label{history}
In this section, we provide a brief overview of the history of Occam's razor and its mathematisation.
The principle of parsimony, most commonly known as Occam’s razor, is named after the medieval monk William of Ockham \cite{sep_ockham}, 
even though it was formulated as far back as Aristotle \cite{charlesworth1955aristotle}.
It is usually translated as:
”entities should not be multiplied without necessity”,
which is to be interpreted to mean that given multiple conflicting theoretical models that equally well explain all of the observations made so far,
one should choose the simpler model, i.e.: the explanation that contains the least assumptions,
or more precisely, the smallest amount of presupposed information.
Note that the simplest model does not need to be the easiest one to discover nor the easiest one to understand by any means. 

%The most famous historical application of Occam's razor is the transition from the many epicycles of geocentric models to the simpler, heliocentric Newton's theory of gravity, which .
Occam's razor was later also restated by Newton in his Principia \cite[\mbox{Book III,} \mbox{Rule 1}]{newton1999principia}.
%Afterwards, the discovery of Bayes' theorem provided some mathematical support for Occam's razor, further discussed in \mbox{Section \ref{main section}}.
Many other successful scientists have repeatedly mentioned simplicity as their guiding principle, e.g., Einstein \cite{einstein2011ultimate}.
Einstein's years of stagnation during the development of general relativity can be attributed to him not having taken Occam's razor seriously at first \cite{norton2000nature}, whereas conversely, his later successful development of general relativity can be attributed to him having embraced Occam's razor by that time \cite{norton2000nature}.
Later, multiple attempts were undertaken to prove Occam’s razor based on Bayes’ theorem alone.
However, these lines of reasoning were ultimately deemed to be insufficient \cite{wolpert1995bayesian}.
This is related to Blumer's contested argument \cite{blumer1987occam} mentioned further below.

In the early days of the electronic computer age, 
the seminal papers by Solomonoff \mbox{\cite{solomonoff1964formal1, solomonoff1964formal2}}, Kolmogorov \cite{kolmogorov1965three}, and Chaitin \cite{chaitin1969length} formalized complexity, simplicity, and Occam's razor.
Several other formalizations of Occam’s razor were developed thereafter, 
which collectively may be referred to as ‘minimum description length' (MDL) principles
\mbox{\cite[Def.\hspace{0.5mm}2.7.24]{hutter2024}}\cite{allison2018coding, nannen2010short, rissanen1978modeling}.
The general idea is to select the shortest formal description or the shortest computer program that generates exclusively all of the observed data.
This means that models correspond to programs. 
The length of the program is measured by its information content,
e.g., by how many bits are required to write the program.
Note that the shortest program’s run-time does not need to be short at all. 

Such a modernized Occam's razor, that now lent itself to a comprehensive mathematical treatment, then opened the door to the possibility of an eventual rigorous proof.
The most useful and most difficult step in this direction must arguably have been Levin's coding theorem 
\cite{levin1974laws}\cite[Thm. 4.3.3]{li2008introduction}.
This theorem is sometimes thought to suffice as a complete proof of Occam's razor.
However, on its own, this theorem is not quite sufficient (See \mbox{Section \ref{main section}}) 
and therefore, further improvements were later added by Hutter 
\cite[p.\hspace{0.5mm}143]{hutter2024}
\cite{hutter2010complete_occamProof}
and others 
\cite[p.\hspace{0.5mm}277\hspace{0.5mm}\&\hspace{0.5mm}287]{li2008introduction}\cite{dingle2018input}.
However, to this day, 
such a proof has never become widely accepted nor well-known, 
as is apparent from Sterkenburg's extensive writings \cite{sterkenburg2018universal}.
Our paper now aims to increase the generality, completeness, clarity, accessibility, and credibility of such a proof of Occam's razor.
Our paper is intended as a contribution to epistemology and metascience rather than machine learning.
Nevertheless, we include the following short paragraph on machine learning, for completeness' sake.

In the context of PAC learning theory \cite{valiant1984theory}, 
a less general mathematical argument for Occam's razor was constructed by Blumer \cite{blumer1987occam}.
This was later refined and connected to Kolmogorov complexity by others \cite{li2003sharpening}, while the validity of the older argument became contested \cite{herrmann2020pac,domingos1999role,pittphilsci24910}.
Be that as it may, in modern machine learning, 
it is standard practice to use a so-called ‘regularization term’ in order to minimize the model complexity \cite{tian2022comprehensive}.
Hence, Occam's razor is a vital component of most modern algorithms, responsible for the many impressive results in the field of artificial intelligence.
While the term ‘model complexity’ is common in machine learning, 
the following section, titled Model Complexity, is not concerned with machine learning models, but instead, with scientific models in general.

\section{Model Complexity}\label{model complexity}
Before formalizing and proving Occam's razor, 
one needs to discuss how exactly the complexity of any scientific model is to be quantified in an objective manner.
Such is the purpose of this section here.
In \mbox{Subsection \ref{models_as_algorithms}},
we explain why scientific models generally correspond to some algorithms or programs.
In our paper, the complexity of a model, the model complexity\footnote{
Various less general notions of model complexity exist, e.g., the VC-dimension \cite{vapnik2015uniform}.
}, 
is the amount of information contained in such a program.
However, such a notion of complexity can be scrutinized in various ways due to its dependence on the choice of the programming language.
Therefore, in \mbox{Subsection \ref{reference formalism section}}, 
we resolve several such problems, including the long-standing concerns about circular reasoning.
We conclude that the widely-known, simple formalism called ‘untyped lambda calculus’ \footnote{
The lambda \textit{calculus} \cite{bernays1936alonzo,barendregt1984lambda,church1985calculi} is unrelated to \textit{calculus}, i.e.: 
It is unrelated to differentiation and integration. Lambda calculus is orders of magnitude simpler than calculus.} 
is well-suited for expressing programs such that their information content will quantify model complexity in an objective manner.

\subsection{Scientific Models as Algorithms}\label{models_as_algorithms}

As science aims for objectivity, reproducibility, and falsifiability,
scientific models \cite{sep_models_science} should be sufficiently unambiguous such that their output behavior may not be manipulated to circumvent falsification.
Therefore, any theoretical scientific model should be translatable into a sequence of clear instructions such that anyone could follow them reliably to calculate the predictions.
Such a sequences of instructions were called algorithms, long before computers existed.
Later, it became established that any such algorithm could be automated by a corresponding Turing machine---this is called the Church-Turing thesis \cite[Thesis 2.3]{hutter2005universal}
\cite{sep_church_turing,turing1936computable,bernays1936alonzo}.
Furthermore, every possible Turing machine can be emulated by a so-called universal Turing machine \cite{turing1936computable}, given a corresponding program.
Hence, the full range of theoretical scientific models is captured via the statement that every possible such model corresponds to some computer program.
Vice versa, every possible program can be thought of as such a model, 
if we assume no further restrictions on what constitutes a scientific model, 
i.e., full generality.
A model/program may be deterministic or stochastic, which we discuss in later, 
in \mbox{Section \ref{main section}}, 
where we will firstly employ all possible, deterministic models to prove Occam's razor and then go on to generalize the proof to all possible stochastic models.

\subsubsection{Models versus Theories}\label{model_vs_theory}
Algorithmic information theorists often assume the two terms, ‘model’ and ‘theory’, to be interchangeable with each other \cite{hutter2013subjective}, 
when discussing Occam's razor.
However, in other fields of science, 
a model is considered a mere constituent of a theory \cite{sep_models_science_modelVsTheory}
and a ‘scientific theory’ is often understood to mean ‘a cluster of models’ \cite{giere1988explaining},
Therefore, when a scientist employs Occam's razor to argue for a novel theory, 
the claim ‘the novel theory is simple!’ may be rejected by other scientists,
while the claim ‘this novel model is simple!’ may have been accepted by other scientists.
Hence, one may want to avoid conflating these terms in practice.

\subsubsection{Physical Scale Models}\label{scale_models}
Besides all of these theoretical models, there also exist scale models made from literal matter, 
e.g., miniature seismic shake table models of bridges and buildings employed for earthquake engineering.
At first sight, such physical scale models may seem to lay wholly outside the scope of our paper, as they are material rather than theoretical.
However, in science, any such scale model will still have to be accompanied by a theoretical model that generalizes the scale model's experimental outcomes to greater scales or smaller scales. 
Such a theoretical model will therefore have to contain some information about the scale model as well as some information about the system under study, e.g., the full-scale bridge to be built. 
Occam's razor applies to this theoretical model, which affects the requirements to be fulfilled by the scale model in order for it to be predictive across scales.
Therefore, even scale models made from matter are affected by Occam's razor, albeit less directly.
However, further discussion of such scale models is indeed outside the scope of this paper and henceforth, mentions of models will be referring to the abstract/immaterial/theoretical scientific models.

%The model of the system under study, e.g., curved spacetime, may be accompanied by a series of additional models that represent the different experiments to be performed on the system under study. Let us combine these models of experiments together into one program that acts as a mathematical function $f$. This function $f$ will take as its input the entire model of the system under study. $f$ will then perform the experiments,  and it will then output the series of experimental outcomes. For the sake of generality, the amount of data that these experimental outcomes consist of may be arbitrarily large or arbitrarily small.

\subsection{Choice of the Reference Formalism}\label{reference formalism section}

As mentioned before, in order to formalize and prove Occam's razor, 
we require an objective mathematical definition of the complexity of any theoretical model.
The complexity of a model may be defined as the number of bits of 
information\footnote{Whether the information content of a program is defined as the program's size, or its Shannon information, or its Kolmogorov complexity will not matter, ultimately. 
See \mbox{Sect. \ref{Core Section}}}
contained in the program that corresponds to the model.
However, note that the same model/algorithm will have arbitrarily differing sizes when translated into arbitrary programming languages, 
which is a threat to objectivity.
%Here: Talk about the many authors that discussed the reference machine.
In this context, Kolmogorov referred to such a ‘programming language’ as the ‘description language’ \cite{kolmogorov1965three}, 
and Tromp later referred to it as the ‘description method’ \cite{tromp2007binary}, 
as his language may be considered to be a minimal model of computation rather than just a language.
Solomonoff instead called it the ‘reference language’ \cite{solomonoff2010algorithmic}.
In this context, choosing a language is equivalent to choosing a ‘universal machine’ \mbox{\cite[S.3.1.2.3]{solomonoff1964formal1},}
later commonly referred to as the ‘reference machine’ \cite{li2008introduction} or the ‘reference computer’ \cite{muller2010stationary}.
But which one of these seven terms shall one use?
The term ‘reference machine’ is the most common one, 
but the ideal term should also encompass programming languages and other formal systems,
which is why we shall henceforth refer to this concept as the ‘reference formalism’.

Here, we provide a brief overview of the history of reference formalisms.
Solomonoff himself originally suggested \mbox{\cite[S.3.1.2.3]{solomonoff1964formal1}}  
that a good reference formalism could be McCarthy's LISP \cite{mccarthy1960recursive}.
Chaitin later utilized LISP as a reference formalism \cite{chaitin1977algorithmic} to obtain concrete error bounds concerning Kolmogorov complexity.
The creation of LISP was inspired by an older and simpler formalism called ‘untyped lambda calculus’ \cite{bernays1936alonzo,barendregt1984lambda,church1985calculi} which had long been known to be Turing-complete. 
This formalism was later argued by Tromp \cite{tromp2007binary} to be superior to LISP, in this context.
He therefore developed the ‘binary lambda calculus’ \cite{tromp2007binary,tromp2023functional} which is purpose-built for being a reference formalism and he then improved over Chaitin's concrete results.
This is the current state of the art.
Note that the binary lambda calculus is relatively simple, 
yet it had never even been widely established why and how reference formalisms should be simple in the first place, 
as we will discuss below.

\subsubsection{Objections due to Circularity}\label{circular}
It has long been suggested that the reference formalism should itself be as natural \cite{hutter2009open} or as simple as possible, 
i.e., the least contrived or least complex.
However, when positing this simplicity of the reference formalism,
philosophers \cite{vallinder2012solomonoff,sterkenburg2018universal,neth2023dilemma}
%Note: his thesis section 3.2.5.7. mentions subjectivity
have repeatedly raised the following two valid concerns:\vspace{1.5mm} 
\\
{\textbf{(i)} Firstly, how could we possibly select the least complex reference formalism if the definition of complexity is itself already contingent on the choice of a reference formalism?}\vspace{1.5mm}
\\
\textbf{(ii)} Secondly, if we postulate that the reference formalism ought to be as simple as possible, 
did we not just invoke Occam's razor and therefore, any proof of Occam's razor derived therefrom would consist in a fallacy of circular reasoning?
\vspace{1.5mm}

One valiant attempt at resolving these problems was undertaken by Müller \cite{muller2010stationary} who tried to define the simplest/best/‘most natural’ reference formalisms by equating them to the universal Turing machines that would most easily and most often be emulated by the largest fraction of all other universal Turing machines, when provided with random programs.
Instead of discovering the best reference formalisms however, he discovered the proof that his approach would never work \cite{muller2010stationary}.

Some of the difficulties of past attempts to resolve these problems may stem from the fact that the two objections, (i) and (ii), bare resemblance to each other and therefore, 
they may have been conflated into being only one and the same single objection.
For us it is therefore important to distinguish between the two of them and to resolve them separately, one after another.

\subsubsection{Resolution of these Objections}\label{circular resolution}
The resolution of the first objection, (i), is that rather than looking to advanced mathematics for help, 
one can rely on common sense to decide which reference formalisms are simpler than others.
This is because the original notions of simplicity and complexity originate from common sense and any novel formal notions thereof would therefore have to match commonsensical certainty regardless, 
because otherwise, these formal notions would have to be named something other than ‘simplicity’ or ‘complexity’.

An example of a much too complex reference formalism would be a modern microprocessor architecture, usually consisting of billions of logic gates. 
Conversely, an example of an extremely simple reference formalism is the aforementioned untyped lambda calculus.
The untyped lambda calculus only requires one kind of computational step, called ‘$\beta$-reduction’, which only consists in cutting and pasting, quickly learnable by any child. 
It is so astonishingly simple that a novice may scarcely believe the fact that it is Turing-complete, i.e., that it is computationally universal \cite{barendregt1984lambda,bernays1936alonzo, turing1936computable}.
The existence of the untyped lambda calculus already resolves \mbox{Objection (i)}, because it is already so simple that any further simplification can be considered irrelevant in practice.
%Note for myself: Count of letters, words, or pages. mereology. reference formalism must be the informal natural language.

The solution to the second objection, (ii), is that even \textit{without} invoking Occam's razor, 
there yet does exist a logical reason for the reference formalism to be simple.
Consider the hypothetical scenario of a reference formalism that is a modern programming language, including libraries with complex subroutines.
In such an environment, certain complex models could be expressed in the form of a short source code whose information content will not be reflective of the true complexity of the model.
Hence, such a complex reference formalism would result in an unreliable measure of complexity.
Such unreliability can be prevented entirely through the use of an extremely simple reference formalism, ideally, as bare-bones as the untyped lambda calculus.
Note that we have not invoked Occam's razor in our argument and therefore \mbox{Objection (ii)}, may be considered resolved as well.
A similar resolution was once also formulated by Rathmanner \cite[pp. 1113--1114]{rathmanner2011philosophical}.
However, the following additional objection to this solution could still be raised:\vspace{1.5mm}
\\ 
\textbf{(iii)} The aforementioned unreliability any single complex reference formalism may instead be avoidable via some alternative, equally legitimate approach that employs a large ensemble of complex reference formalisms that all compensate each others flaws, 
which ultimately may lead to equally legitimate results that contradict starting from a simple reference formalism.\vspace{1.5mm}

There would be multiple ways to counter this objection, (iii), but the following conclusion by Müller, who had already pursued such an alternative approach, should suffice \cite{muller2010stationary}: 
"There is no way to get completely rid of machine-dependence, neither in [his] approach nor in any other similar but different approach."
This means that any such effort will self-defeat because the initial choice of a (simple) machine or reference formalism will turn out to be a prerequisite for any such approach that is based on many complex reference formalisms.
Thus, Objections (iii) and (ii) are now resolved.
%Secondly, even if, hypothetically, such a sophisticated alternative approach were to succeed, 
%it would not matter as one can expect it to lead to the same experimental predictions because our ‘election’ of predictions in \mbox{Section \ref{Core Section}} already starts from us charitably favoring all the arbitrarily complex models.
%An objection to this would be that even if the predictions are identical, occams razor might still turn out wrong because the simplest models would not be simple according to the alternative. 
%However, rather than satisfying all possible definitions of simplicity, what matters is that there exists some reasonable definition of simplicity, for which Occam's razor can be proven under reasonable assumptions.

\subsubsection{Discussion of the Lambda Calculus}\label{lambda}
According to our two previous subsubsections, 
the untyped lambda calculus (ULC) would be well-suited to be our reference formalism.
However, novices may raise the following objection:\vspace{1.5mm}
\\
\textbf{(iv)} The ULC may turn out to be such an inconcise language for formulating algorithms that highly complicated algorithms may require multiple times more bits in order for them to be formulated in the ULC than if the same complicated algorithms were to be formulated in terms of some alternative, similarly simple reference formalism.\vspace{1.5mm}

This worry is resolved by understanding that the ULC can be used to emulate any other simple reference formalism via a relatively small program, if need be, 
and thus, a complicated algorithm could be formulated in ULC as a combination of the small emulator program plus the complicated algorithm formulated in terms of the alternative reference formalism.
Therefore, the deviations between these two alternative measures of model complexity can be limited to the size of the small emulator program, i.e.:
The deviations will remain small relative to the size of complicated algorithms and thus, 
the complicated algorithms cannot require multiple times more bits when formulated in ULC than in the alternative reference formalism.
Hence this objection, (iv), is resolved.
This insight originates from the proof of the invariance theorem \mbox{\cite[S.3.1.2.3]{solomonoff1964formal1}}\cite{kolmogorov1965three,chaitin1969length,li2008introduction, tromp2007binary} whose discovery is viewed as the birth of the field of algorithmic information theory.
With these four objections, (i), (ii), (iii), and (iv), resolved, we will henceforth consider the ULC to be well-suited as a reference formalism.
However, one final objection may be raised against the ULC:\vspace{1.5mm}
\\
\textbf{(v)} Since all of the algorithmic information theory literature was formulated in terms of binary coding, it may be incompatible with the ULC, which is not based on binary coding.\vspace{1.5mm}

Out of multiple ways to resolve this objection, (v), the easiest way, currently, is for us to point to Tromp's purpose-built \textit{binary} lambda calculus (BLC) \cite{tromp2007binary,tromp2023functional}.
The BLC, while not quite as simple as the ULC, 
only "adds the minimum amount of machinery to enable binary input and output" \cite{tromp_2012}.
Since our proof of Occam's razor is based on traditional algorithmic information theory, which is based on binary coding, 
all of the following sections may be thought of as employing the BLC as their reference formalism.
However, we will rely on the older standard notation \cite{li2008introduction} and a closer study of the BLC is currently unnecessary for the study of just our paper.

Outside of traditional algorithmic information theory, 
the BLC may not always be the optimal reference formalism to be used, since in principle, \textit{binary} I/O is not strictly required for modeling nor computation.
In that case, one may want to use a different ULC-based system to quantify model complexity, 
whose inception is outside the scope of our current paper.
In any case, our general proof of Occam's razor would also translate onto any reasonable, alternative, future, ULC-based system,
intended for the quantification of model complexity.

\section{General Mathematical Proof}\label{main section}
This section contains the main argument for Occam’s razor, including our proofs of the supporting mathematical theorems.
We start by introducing the widely-known Kolmogorov complexity along with some relevant theorems about it, in \mbox{Subsection \ref{K_section}}.
The theoretical field that is concerned with Kolmogorov complexity is called ‘algorithmic information theory’ (AIT).
In \mbox{Subsection \ref{weak_section}}, we review \textit{weak} AIT-based arguments for Occam’s razor, 
before moving on to \mbox{Subsection \ref{Core Section}}, where we present the \textit{strong} argument for Occam's razor, that is also based on AIT.
Since theorems contain seemingly problematic deviations,
in \mbox{Subsection \ref{deviations section}}, 
we resolve two objections concerning the deviations and one objection concerning the incomputability of Kolmogorov complexity.
In \mbox{Subsection \ref{stochastic section}}, we generalize our proof technique from deterministic to stochastic models.
Finally, in \mbox{Subsection \ref{assumptions}}, we gather the five reasonable initial assumptions that we have relied on throughout our paper.

\subsection{Prefix-Free Kolmogorov Complexity}\label{K_section}
In this subsection, we will introduce the widely-known Kolmogorov complexity along with some simple theorems about it,
on which our proof in \mbox{Subsection \ref{Core Section}} will rely.
Kolmogorov complexity \cite{hutter2008scholarpedia,li2008introduction}, also known as ‘algorithmic complexity’ or ‘program size complexity’, is a general measure of information content that is so highly general that it has lead to the mathematisation of a wide range of general concepts,
such as randomness \cite{martin1966definition,downey2010algorithmic}, 
similarity \cite{li2004similarity,vitanyi2005universal}, 
intelligence \cite{legg2006formal,hutter2024}, 
thermodynamic entropy \cite{zurek1989algorithmic,ebtekar2025foundations}, 
and so forth.
The Kolmogorov complexity of some data $x$, often denoted as $K(x)$,
is defined as the size of the smallest possible program that will compute and output $x$.
For novices, the following may serve as quick food for thought:
Let $|x|$ be the size of $x$, i.e., how many bits $x$ consists of. 
Then, $K(x)\approx|x|$ would indicate that $x$ is nearly completely random, while, on the other hand,
$K(x)\ll|x|$ would indicate that $x$ is quite nonrandom and compressible.
This concludes our informal introduction of $K$ and we now move on to a more specific definition of $K$.

There exist a few different versions of Kolmogorov complexity.
For this paper, we will only be using the widely-used, so-called ‘prefix-free’ Kolmogorov complexity, also known as (algorithmic) ‘prefix complexity’ \cite{hutter2008scholarpedia, levin1974laws, chaitin1975theory}\cite[Chap. 3]{li2008introduction},
which was later also concretized by Tromp \cite[Def. 4]{tromp2007binary,tromp2023functional}.
Given some object $x$, which is usually a binary sequence,
its prefix-free Kolmogorov complexity $K(x)$ is the size of the smallest possible, self-delimiting program $p$ that will have returned only $x$ upon halting.
This is commonly denoted as $U(p)=x$, 
where $U$ stands for ‘universal’,
which represents the reference machine, 
or rather, the refernece formalism
(\mbox{Subsect. \ref{reference formalism section}}).

For the ‘conditional’ prefix-free Kolmogorov complexity $K(x|y)$,
some extra information $y$ may be provided, 
ready to be processed by $p$.
This may usually be denoted as $U(y,p)=x$.
The more that $y$ contains information helpful to computing $x$, the less such information will be required to be contained within $p$,
and hence, the smaller $K(x|y)$ will be.
This means that $K(x|y)$ is a a general measure of how informative $x$ is, 
conditioned on $y$ already having been known.
Let $\epsilon$ denote a sequence of length zero, i.e., an empty string.
Then, we can connect the unconditional notion with the conditional notion via the definitions $U(p):=U(\epsilon,p)$ and 
$K(x):=K(x|\epsilon)$.
All together, this may usually, formally be denoted as follows:

\begin{definition}\label{K Def}
The (Conditional) Prefix-Free Kolmogorov Complexity ($K$)
\[K(x|y)\ :=\ \min_{p\in\{0,1\}^*}\ \{|p|:U(y,p)=x\}
\ \ \ \ \ ,\ \ \ \ \ \ \ 
K(x)\ :=\ K(x|\epsilon)\],
where $|p|$ is just the size of the program, 
i.e., the number of bits that $p$ consists of;\
$U$ is a function that represents the reference formalism (\mbox{Subsect. \ref{reference formalism section}});
$y$ is some supplementary data that is provided to $U$ as an auxiliary input, such that $y$ may be read and processed according to a program $p$. \
\mbox{$U(y,p)=x$ } means that $U$ interprets and executes a self-delimiting program $p$, that results in returning only $x$ upon halting.
A self-delimiting program $p$ is a program that causes $U$ to read\footnote{
The details of how $U$ reads (or ignores) bits are not important for our paper. 
These details were originally formulated based on ‘prefix Turing machines’ \cite[Chap. 3]{li2008introduction} 
and then later reformulated by Tromp \cite[Def. 4]{tromp2007binary,tromp2023functional} without the use of Turing machines.
Other ULC-based approaches may also be possible in the future.
}
all of the bits of $p$, but no bits, that may had seamlessly been appended to the end of $p$.\vspace{1mm}
\end{definition}

In this sense, self-delimiting programs are already inscribed with the information about their own size, thus making \textit{prefix-free} Kolmogorov complexity a more complete measure of information than the original, \textit{plain} Kolmogorov complexity, which was not based on self-delimitation  \cite{li2008introduction,kolmogorov1968}.
The reason for the name ‘prefix-free’ is that, logically, no self-delimiting program could ever constitute the first part (prefix) of another self-delimiting program, as it would lead to a contradiction about where $U$ should stop reading bits.
Due to this prefix-free property, the famous Kraft–McMillan inequality \cite{kraft1949device, mcmillan1956two} applies as follows:

\begin{theorem}\label{Kraft_conditional}
The Kraft Inequality for conditional prefix-free $K$
    \[
    \forall y:\ \sum_{x}\ 2^{-K(x|y)}\ \ \leq\ 1
    \]
\end{theorem}
\begin{proof}
Given any fixed $y$, for each $x$ there must exist a different, minimal program $\chi$ with $U(y,\chi)\ =x$. 
All of these distinct, self-delimiting programs $\chi$ together constitute a so-called ‘prefix code’
which just means that no self-delimiting programs will be a prefix of another self-delimiting program, as mentioned earlier.
When there is a prefix code, 
the Kraft–McMillan inequality \cite{kraft1949device, mcmillan1956two} applies.\\
q.e.d.
\end{proof}

Throughout the history of algorithmic information theory, \mbox{Theorem \ref{Kraft_conditional}} has often come in handy when proving more advanced theorems.
The same is true in our paper.
Before we state the next theorem,
let us recall an analogous, older, and simpler theorem from Shannon's classic information theory, 
namely, the chain rule of conditional entropy for two random variables $X$ and $Y$ \cite{thomas2006elements,shannon1948mathematical}:
\begin{align}\label{Shannon Chain}
H(X,Y) = H(X)+H(Y|X)
\end{align}

, where $H(X,Y)$ is the joint entropy and $H(Y|X)$ is the conditional entropy.
The proof is short and simple and can be found in standard texts \cite{thomas2006elements,shannon1948mathematical}.
Since the conditional $H$ and the conditional $K$ are conceptually similar to each other,
the question naturally arises whether such a chain rule also exists for $K$.
In order to formulate such a rule, the joint $K$ needs to be defined first, 
i.e., the joint prefix-free Kolmogorov complexity $K(x,y)$.
It could easily be defined a via concatenation $xy$ of $x$ and $y$, 
i.e., append $y$ to $x$, and define $K(x,y):=K(xy)$.
However, note that the concatenation does not carry the information of the size $x$ nor the size of $y$ due to the missing demarcation of where $x$ ends and $y$ starts.
Therefore, in order not to lose any such information,
the joint prefix-free Kolmogorov complexity is usually defined as $K(x,y):=K(\langle x,y\rangle)$,
where the data $\langle x,y\rangle$ is defined to be similar to a concatenation, except that it contains the extra bits that provide the information that can be used to demarcate where $x$ ends and $y$ begins.
We can call $\langle x,y\rangle$ an ‘encoded pair’.
There exist various ways to define $\langle x,y\rangle$ in more detail\footnote{
The simplest encoding of pairs/sequences is in terms of Church pairs in ULC/BLC
\cite[Sect. 2.1]{tromp2023functional}. 
}, 
but these details are not important now.
In analogy to the properties of the joint $H$, for the joint $K$, 
the following three basic properties hold, up to an error term $\pm\mathcal{O}(1)$ bounded by some small constant:
\begin{align}\label{joint K properties}
K(x,y)=K(y,x)\ \ ,\ \ \ \
K(x,y)\geq K(x)\ \ ,\ \ \ \
K(x,y) \leq K(x)+K(y)
\end{align}
These equations relatively easy to prove.
As a tip:
For the first equation, the error term is bounded by a constant equal to the size of some small algorithm that just converts $\langle x,y\rangle$ into $\langle y,x\rangle$.
Note that this constant is independent of $x$ and solely determined by which reference formalism (Sec. \ref{reference formalism section}) one had selected initially.

Now that we have introduced the joint $K$,
we can state a chain rule for $K$ analogous to Shannon's aforementioned chain rule of conditional entropy (Eq. \ref{Shannon Chain}):\\
\mbox{$K(x,y) \approx K(x)+K(y|x)$}. 
However, it is known that the error of this equation is not bounded by a constant \cite[p.\hspace{0.5mm}244-247]{li2008introduction}, 
i.e.: The difference between the two sides is not in $\pm \mathcal{O}(1)$. 
Thankfully, via a small modification of the equation, namely, 
by replacing the term $K(y|x)$ with $K(y|x,K(x))$, a higher accuracy can be achieved, 
i.e.: The error will be then be bounded by a constant.
Thereby, the meaning of $K(y|x,K(x))$ is that, in addition to passing just $x$ as auxiliary information, 
also a binary representation of the integer $K(x)$ is passed as auxiliary information. 
This is only a small addition of information since, generally, 
the information content of any integer will be much smaller than the integer itself, as it only grows logarithmically.
This technical detail is not important, other than the fact that it ensures the error will always remain in $\pm \mathcal{O}(1)$.
This theorem can be stated as follows:

\begin{theorem}\label{chain rule theorem}
The Chain Rule for Conditional Prefix-Free Kolmogorov Complexity\\
\[\forall x,y: \ \
K(x, y)\ =\ K(x)\ +\ K(y| \ x, K(x)) \ 
\pm \mathcal{O}(1)\]
\end{theorem}

\begin{proof}
The proof of this equation can be found in the standard text \mbox{\cite[Thm. 3.9.1]{li2008introduction}}, 
even though they did not call it ‘the chain rule’. 
q.e.d.
\end{proof}

We have studied this old and complex proof \mbox{\cite[Thm. 3.9.1]{li2008introduction}} in detail and we can therefore attest to its veracity.
However, most readers may want to avoid reading this time-consuming proof of the chain rule.
In order to, nevertheless, develop a good sense of why this chain rule is true,
one may want to study the simpler proof \cite[\mbox{Thms. 20, 21, 22}]{shen2017kolmogorov}
of the chain rule for the \textit{plain} Kolmogorov complexity in its stead.
Note however, that in turn, this chain rule for the \textit{plain} Kolmogorov complexity carries a logarithmic error term, rather than the constant error term $\pm\mathcal{O}(1)$.
Due to this logarithmic error term, 
this cheaper version of the chain rule is not quite precise enough to support the proof of our \mbox{Theorem \ref{Core}},
yet, this version may satisfy the reader.
If this is still too complex, 
the reader may intuit \mbox{Theorem \ref{chain rule theorem}} by analogizing to Shannon's aforementioned chain rule of conditional entropy (Eq. \ref{Shannon Chain}).

In addition, we will also make use of slightly more general version \mbox{(Corr. \ref{chain rule corollary})} of the chain rule, analogous to the following equation,
which is identical to Shannon's chain rule \mbox{(Eq. \ref{Shannon Chain}),} 
except that every term is now conditioned with some additional random \mbox{variable $Z$:}
\begin{align}\label{Shannon Chain general}
H(X,Y|Z)\ =\ H(X|Z)\ +\ H(Y|X,Z)    
\end{align}
Note that this simple equation about $H$ is quite similar in form and meaning to the following corollary about $K$:
\begin{corollary}\label{chain rule corollary}
a Generalized Chain Rule for Conditional Prefix-Free $K$
\[\forall x,y,z: \ \
K(x,y|z)\ =\ K(x|z)\ +\ K(y|\ x,K(x|z),z)\ 
\pm \mathcal{O}(1)\]    
\end{corollary}
\begin{proof}
The proof is analogous to the proof of \mbox{Theorem \ref{chain rule theorem}}.
This corollary is also mentioned in the standard text by Li \& Vitanyi
\cite[Equation 3.22]{li2008introduction}.
q.e.d.
\end{proof}

\subsection{Weaker Arguments for Occam's Razor}\label{weak_section}
The purpose of this subsection is to review similar but weaker arguments for Occam's razor 
in order to prevent them from being conflated with the strong argument to be introduced in the next subsection (\ref{Core Section}).
If this is not of interest to the reader, they are invited to skip forward directly to \mbox{Subsection \ref{Core Section}}.

We have already mentioned some weak attempts at proving Occam's razor in \mbox{Section \ref{history}}, 
such as the ones based on Bayes theorem alone, 
that were refuted by Wolpert \cite{wolpert1995bayesian}.
Per se, there is nothing wrong with employing Bayes' theorem to argue for Occam's razor. 
However, the lack of the concept of Kolmogorov complexity is what made the past Bayesian arguments fail, 
as they lacked the aforementioned generality (\ref{K_section}) that Kolmogorov complexity would have enhanced them with.
Kolmogorov complexity was being combined with Bayes theorem right from its inception by Solomonoff \cite{solomonoff1964formal1} in order to formalize the correct application of Occam's razor, rather than proving it. 
While there is nothing wrong with Bayes' theorem, 
we will not be using Bayes' theorem in our paper simply because it is not necessary.

Sterkenburg, who had written extensively on the potential drawbacks \cite{sterkenburg2018universal,neth2023dilemma} of algorithmic probability theory,
has identified one insufficient argument 
\cite[p. 122]{sterkenburg2018universal} \cite{sterkenburg2016solomonoff} for Occam's razor.
This argument was derived from Solomonoff's convergence theorem \cite{solomonoff1978complexity}, a theorem which demonstrated the reliability of Solomonoff's universal prediction scheme \cite{solomonoff1964formal1,solomonoff1964formal2}.
While the theorem is in itself correct, 
Sterkenburg refuted the therefrom derived argument for Occam's razor on grounds of implicit circular reasoning \cite[p. 124]{sterkenburg2018universal} \cite{sterkenburg2016solomonoff}.
Since he already concluded this issue, let us now move on to another weak argument for Occam's razor.

A stronger argument \cite{standish2004occam}, although not strong enough,
consists in arguing directly from Levin's coding theorem  
\cite{levin1974laws}\cite[Thm. 4.3.3]{li2008introduction} in its raw form. 
We dedicate the rest of this subsection to this argument.
The idea of this argument is to start in a neutral, unbiased fashion, i.e., 
from a uniform probability distribution over all possible binary sequences of some length $n$.
A binary sequence can be sampled via a sequence of $n$ fair coin flips, each of which has a 50:50 chance to result in a zero or a one.
Let $n$ go towards infinity.
If such a binary sequence is provided to a computer and executed as a program, the computer may produce some output and halt. 
Given such a randomly sampled program, some lay people may initially assume that all possible outputs will therefore have an equal probability of occurring as each other.
However, this is not the case because there is no one-to-one-correspondence between programs and outputs.
The probability that, upon halting, data $x$ will be the output, is commonly denoted as $m(x)$ and it is called the ‘discrete universal a priori probability’ of $x$ \cite{Scholarpedia_2007}.
Remarkably, there exists a close relationship between this $m(x)$ and $K(x)$ (\mbox{Def. \ref{K Def})}.
This relationship is known as Levin's ‘Coding Theorem’
\cite{levin1974laws}\cite[Thm. 4.3.3]{li2008introduction} and it states that, for any $x$, one can use $K(x)$ in order to obtain a range within which $m(x)$ has to lie, as follows:
\begin{align}\label{coding theorem}
2^{-K(x)}\ \leq\ m(x)\ <\ 2^{-K(x)+\mathcal{O}(1)}
\end{align}
, where $\mathcal{O}(1)$ is again just a deviation that can never exceed a particular, constant value that is independent of $x$ and solely determined by which reference formalism (Sec. \ref{reference formalism section}) one had selected initially.
Levin's Coding Theorem was proven in a sophisticated manner that we will not outline here.
As is evident from this formula, a priori, any observation/output $x$ is exponentially less likely to occur the more complicated the simplest program is that would return $x$, 
i.e.: A natural bias towards simplicity seems to have emerged, even-though these probabilities, $m(x)$, stem from \textit{long} bit sequences ($n\rightarrow\infty$) and our only initial assumption was the indifference between the long sequences, 
i.e., the principle of indifference \cite{keynes1921chapter}. 
Therefore, this line of reasoning could already be considered a proof of Occam's razor. 
However, we will now go on to raise an objection to this potential proof.

Recall from \mbox{Subsection \ref{K_section}} that we only considered the lengths of special kinds of prefixes $p$ that are self-delimiting programs such that whatever information may be appended to $p$ will be ignored by the computation as the computation will already have returned an output $x$ and halted before reading anything beyond the end of $p$.
For the aforedescribed uniform distribution, the probability that a long bit sequence will start with any given, shorter sequence $p$, is obviously equal to $2^{-|p|}$,
where $|p|$ is the length of the prefix $p$.
Therefore, the aforementioned discrete universal a priori probability, $m(x)$, can simply be written as follows, as it is also commonly denoted:
\begin{align}\label{universal prior}
m(x)\ =\ \sum_{p:U(p)=x}2^{-|p|}
\end{align}
In this sum, smaller programs $p$ can be seen to be weighed much more heavily than longer programs $p$, despite us having started from a uniform distribution.
Considering this bias towards simplicity, 
one may now object by arguing as follows:\vspace{1.7mm}
\\
\textbf{(vi)} Since only the prefixes $p$ are relevant for computation, only these prefixes may be considered programs and models, rather than all the redundant, longer bit sequences. 
Therefore, the previous application of the principle of indifference (over redundant bit sequences) may have been invalid and the principle of indifference should instead have been applied over these prefixes $p$, upon which the bias toward simplicity may disappear, thus potentially rendering this argument for Occam's razor invalid.\vspace{1.7mm}

One attempt at rebutting this objection, 
would be to argue that, for obvious combinatorial reasons, there exists a much larger number of long programs than simple programs, which did already neutralize this ‘unfair’ bias towards simplicity in our original sum, hence, Occam's razor would remain justified.
However, we can reject this rebuttal for the following reason.
Since $m(x)$ is a probability, we know this sum \mbox{(Eq. \ref{universal prior})} has to converge to a finite value, which implies that, in this sum, the total contribution of increasingly longer programs $p$ has to go toward zero.
This means that the increasing quantity of increasingly longer programs is insufficient to neutralize the potentially ‘unfair’ bias toward simplicity.
Hence, \mbox{Objection (vi)} remains a valid concern.
The therefore weakened argument for Occam's razor nevertheless remains valuable in that it serves as a direct predecessor to the improved argument laid out in the following subsection (\ref{Core Section}),
which is then able to withstand \mbox{Objection (vi)} because a fair bias toward simplicity prevails despite the principle of indifference then indeed having been applied over sets of self-delimiting programs $p$, rather than over redundant bit sequences.
Historically, Levin's aforementioned ‘Coding Theorem’ has also served as a basis for proving the chain rule \mbox{(Thm. \ref{chain rule theorem})}.
We make use of this chain rule in the following subsection.

\subsection{Democracy of Deterministic Models}\label{Core Section}
The main argument, which should prove a modernized Occam's razor beyond a reasonable doubt, 
is that if we consider the set of all possible, valid, theoretical scientific models \mbox{(Sec. \ref{model complexity})} of an arbitrarily large size $n$ (model complexity $n\to\infty$), 
that are consistent with our past observations $o$,
then, the future observations predicted by these models will democratically agree with the predictions of the objectively smallest possible models consistent with $o$.
So, despite us having provided all of this charitability toward arbitrarily high model complexities, the outcome will still end up agreeing with Occam's razor.

All these complex models of size $n$ are assigned equal prior probabilities of being right\footnote{
The proportion/fraction/frequency of models that predict some $x$ then determines the probability of $x$,
which one could therefore interpret to be a frequentist probability \cite{mood1950introduction,venn1866logic}.
}, 
which may also be described as a ‘fair democracy’,
i.e., fair between such equally complex models at least (small models are excluded). 
This equiprobability stems from the principle of insufficient reason, also known as the principle of indifference \cite{keynes1921chapter}, which we have also mentioned in the previous subsection.
The main difference to the previous subsection is that, this time around, this set consists of only models that are \textit{self-delimiting} programs rather than it consisting of all possible binary sequences.
The reason for this upgrade is \mbox{Objection (iv)},
since, this objection is resolved if, 
despite only accepting the votes of self-delimiting programs (prefixes $p$),
the outcome will still agree with the simplest models,
as we will go on to prove mathematically in what follows.

%Paragraph giving credit
But first, let us briefly mention the preceding literature on this matter.
Such a justification of Occam's razor,
although less comprehensive, can also be found in the standard text 
\mbox{\cite[Example 4.3.5 \& Exercise 4.3.6]{li2008introduction}},
which is cited by Hutter \mbox{\cite[Sec. 8]{hutter2010complete_occamProof}}, 
who relates this argument to multiverse theories and
who also credits Schmidhuber \cite{schmidhuber2000algorithmic}.
Similar techniques were later employed in the interdisciplinary paper by \mbox{Dingle et al.} \cite{dingle2018input} 
in order to explain why various complex systems, 
ranging from biochemistry to financial markets, can be understood via simpler models.
Although none of these earlier works have mentioned \mbox{Objection (iv)} explicitly,
it must have been on the minds of these authors
as none of them relied on the afore-discussed weaker argument \cite{standish2004occam} to which \mbox{Objection (iv)} applies.

Despite some of this technical literature having existed for decades, 
such a mathematical proof of Occam's razor had never become widely known among philosophers, 
because most of this previous literature on proving Occam's razor had been too scattered over disconnected papers and chapters, 
while leaving various critical questions unanswered, 
while invoking superfluously strong initial assumptions \mbox{(\ref{superfluous})},
and furthermore conveyed no easily digestible, mathematical understanding of why the outcome of the democracy should agree with simple models. 
We aim to change this state of affairs by providing resolutions to the twelve objections spread throughout our paper, 
and by listing sufficiently weak initial assumptions \mbox{(\ref{our own assumptions})} for them to be accepted as true by most intellectuals,
and by providing the following, 
relatively easily digestible\footnote{
Our proof does not mention semi-measures, nor Shannon-Fano codes, nor dovetailing.
Such sophistication is tucked away into the proof of the chain rule (Thm. \ref{chain rule theorem} \& Corr. \ref{chain rule corollary}).
} 
mathematical proof, 
that should lend itself to a better intuitive understanding, based on the afore-described chain rule.

We will now start with only deterministic models and we will generalize these results to stochastic models in \mbox{Subsection \ref{stochastic section}}.
Optionally, in order to increase the realism and generality, we can now introduce the so-called ‘independent variables’ $z$.
The adjustable settings of an experiment are what scientists generally call ‘independent variables’, 
which can affect the outcome (dependent variables) of the experiment.
An experimenter may run the same kind of experiment multiple times while adjusting the independent variables, in order to observe how these changes affect the outcomes.
We can represent these optional, independent variables as a binary code $z$.
in principle, one could even use $z$ to encode a series of different experiments, including novel or hypothetical experiments.
If, on the other hand, there are no independent variables at all, one can set $z$ equal to $\epsilon$,
i.e., an empty string.

In order to derive an equation between the predictions of the simplest models and the outcome of the ‘democratic election’, 
we first need to, for each candidate future observation (e.g., $a$ vs $b$),
define the set of all of its ‘voters’, 
each of which is a deterministic model, which is a self-delimiting program of length $n$, consistent with past observations $o$, if it is provided with the independent variables $z$.
We can say that such a model is consistent with past observations $o$ if it returns $o$, and then, we can say that it predicts some future observation $a$ if it then goes on to return $a$ after $o$, such that the complete output will be the concatenation $oa$, 
which would also mean that this model ‘voted’ for $a$.
In the following definition,
we will denote such complete outputs, such as $oa$, by just one letter $x$, which will better match most theorems in this paper.

\begin{definition}\label{voteSetDef}
Let $\mathcal{V}_n(x|z)$ be the set of all self-delimiting programs $p$ of size $n$ that will, if provided with an auxiliary input $z$,
return only the output $x$ and then halt,
i.e.:
\[\mathcal{V}_n(x|z)\ :=\ \{p :\ |p|=n \ \land \ U(z,p)=x\}\]
Where $U(z,p)=x$ carries the same meaning as in \mbox{Definition \ref{K Def}}.
\end{definition}

The fact that that $\mathcal{V}_n(x|z)$ is a subset of all binary codes of length $n$, immediately provides us with an obvious upper bound on the size of this set: $|\mathcal{V}_n(x|z)| \leq 2^n$ .
A much more accurate upper bound will be provided later in this subsection (\mbox{Theorem \ref{Core}}).
Recall the Kolmogorov complexity $K$.
The size of the smallest possible model that predicts $a$ or $b$ can now be written as $K(oa|z)$ or $K(ob|z)$, respectively.
Therefore, according to what we have claimed above, about the outcome of the democracy, 
if $K(oa|z)$ is smaller than $K(ob|z)$,
we can expect $|\mathcal{V}_n(oa|z)|$ to be greater than $|\mathcal{V}_n(ob|z)|$, and vice versa.
In fact, more precisely, as we will go on to prove in this subsection, 
the relationship between these quantities is \textit{exponential}, as illustrated by the power of 2 at the end of the following equation:
\begin{align}\label{odds exp equation}
\text{probability ratio }a\text{ vs }b
\ :=\ \frac{P(oa|z)}{P(ob|z)}
\ =\ \frac{ |\mathcal{V}_n(oa|z)|}{ |\mathcal{V}_n(ob|z)|}
\ \approx\ 2^{ K(ob|z)-K(oa|z)}
\end{align}
So for example, if the smallest model that predicts $a$ is just 30 bits of information larger than the smallest model 
that predicts $b$,
then it will be $2^{30}\approx$ one billion times more likely that $b$ is the correct prediction.
The fact that such a small difference in information can cause such an extremely one-sided probability ratio is what underpins the strength of the democracy-based argument above. 
These numbers should also demonstrate the great utility of a modern Occam's razor as a guiding principle for determining what is true or false in the theoretical sub-fields of science.

For the remainder of this subsection,
we will be concerned with proving \mbox{Equation \ref{odds exp equation}} above,
or, as stated in greater detail, \mbox{Corollary \ref{odds exp corollary}} and \mbox{Equation \ref{odds computable}}, further below.
In the buildup toward that, 
we start by proving a lower bound on the number of votes $|\mathcal{V}_n(x|z)|$,
for any arbitrarily large $n$, as long as $n$ is not much too small:
\begin{theorem}\label{coreEasyDir}
\textbf{Lower Bound on $|\mathcal{V}_n(x|z)|$ :}
\\
For any $n \geq K(x|z)+K(n)+\mathcal{O}(1)$,
the following inequality will hold:
\[|\mathcal{V}_n(x|z)| \ \ \geq \ \ 2^{n-K(x|z)-K(n) - \mathcal{O}(1)} \]
\end{theorem}

\begin{proof}
Let $\chi$ be the smallest possible self-delimiting program that returns only $x$, given $z$, 
and let $\eta$ be the smallest possible programs that returns only $n$.
So, obviously, $|\chi|=K(x|z)$ and $|\eta|=K(n)$.
Let $r\chi\eta g$ be the concatenation of  $r$, $\chi$, $\eta$, and $g$, where $g$ is of such a size that $|r\chi\eta g| = n$.
Thereby, let $r$ be the smallest possible algorithm, purposed such that, if \textit{any} two concatenated, self-delimiting programs $\chi$ and $\eta$ are appended to $r$,
then $r$ will cause both, $\chi$ and $\eta$, to be read and executed,
where an auxiliary input $z$ can be accessed by $\chi$ but not by $\eta$,
and where $n$, the result of $\eta$, will not be returned but will instead be utilized to instruct $U$ to read the $n$ bits of the aforementioned concatenation, 
but no further appended bits,
while the result of $\chi$ will have been returned upon halting, and nothing further,
i.e., $U(y,r\chi\eta g) = x$.
Therefore, $r\chi\eta g\in\mathcal{V}_n(x|z)$.
Note that $r$ is independent of the specific $y$, $\chi$, $\eta$, or $g$, and therefore, $|r|$ is a constant,
i.e., $|r|=\mathcal{O}(1)$.
Furthermore, note that the output will be $x$, regardless of the content of $g$ which means that $g$ could be equal to any of $2^{|g|}$ possible binary codes,
and thus, there exist at least $2^{|g|}$ different programs of the form $r\chi\eta g$ within $\mathcal{V}_n(x|z)$.
Hence, $|\mathcal{V}_n(x|z)| \geq 2^{|g|}$.
Recall that $|r\chi\eta g| = n$.
We can write this as $|r\chi\eta|+|g| = n$,
and therefore $|g|=n-|r\chi\eta|$.
Combined with the previous fact, namely $|\mathcal{V}_n(x|z)| \geq 2^{|g|}$,
we obtain the following: $|\mathcal{V}_n(x|z)| 
\geq 2^{n-|r\chi\eta|} $
$= 2^{n-|\chi|-|\eta|-|r|}$.
If we combine this inequality with the three aforementioned identities $|\chi|=K(x|z)$, $|\eta|=K(n)$,
and $|r|=\mathcal{O}(1)$,
we obtain $|\mathcal{V}_n(x|z)|$
$= 2^{n-K(x|z)-K(n)-\mathcal{O}(1)}$
\\q.e.d.
\end{proof}

Now, that we have proven a tight lower bound ($\geq$) on the number of votes $|\mathcal{V}_n(x|z)|$, 
we will go on to prove an equally tight upper bound ($\leq$) on it.
This will be \mbox{Theorem \ref{Core}},
but in preparation for this proof,
it will be more practical if we first derive upper and lower bounds on the Kolmogorov complexities $K$ of the voters $p$ themselves:

\begin{lemma}
\label{nontrivial lemma}
\textbf{Bounds on $K(p)$}
\\
For any $p \in \mathcal{V}_n(x|z)$, the following inequalities will hold:
\[n+\mathcal{O}(1)\ \ \geq\ \ \ K(p)\ \
\geq\ \ \ K(n)\ +\ K(x| z')\ +\ K(p| z',K(z'))\
-\mathcal{O}(1)\],
where $z':=\langle z, \langle n,K(n)\rangle\rangle$ .

\end{lemma}
\begin{proof}
Let us first show the trivial inequality
$n+\mathcal{O}(1)\geq K(p)$.
Recall that $|p|=n$ (\mbox{Def. \ref{voteSetDef}}),
so we just need to show
$|p|+\mathcal{O}(1)\geq K(p)$.
It suffices to consider an algorithm of constant size ${O}(1)$, 
which will return a copy of any self-delimiting program $p$ that may have been be appended to it,
i.e. it acts as a print command.
As mentioned earlier, a self-delimiting program $p$ is already inscribed with the information about its own size and therefore, 
the ‘print command’ requires no further information besides $p$.
This ‘print command’ combined with $p$, forms a program of size $|p|+\mathcal{O}(1)$ that returns $p$.
Therefore, the size $K(p)$ of the smallest possible program that returns $p$ can be no greater than $|p|+\mathcal{O}(1)$ \ \mbox{(q.e.d.)}.

For the second inequality
$K(p)\ \geq\ K(n)+K(x| z')+K(p| z',K(z'))-\mathcal{O}(1)$ ,
the following is its proof as one string of mostly self-evident inequalities, with corresponding, supplementary, explanatory comments to the right:\\
$K(p)$ \
$\geq\ K(n,p) -\mathcal{O}(1)$
\hspace*{\fill} (constant-sized algorithm extracts $n=|p|$ from $p$)\\
$\geq\ K(n)+K(p|\ n,K(n)) -\mathcal{O}(1)$
\hspace*{\fill} (via the Chain Rule for $K$, Thm. \ref{chain rule theorem})
\\
$\geq\ K(n)+K(p|\ z,n,K(n)) -\mathcal{O}(1)$
\hspace*{\fill} ($z$ lowers $K$, if false, skip $z$: Still, size $=\mathcal{O}(1)$)\\
$\geq\ K(n)+K(x,p|\ z,n,K(n)) -\mathcal{O}(1)$
\hspace*{\fill} (const.-sized algor. to compute $\langle x,p\rangle$ from $p$)\\
$\geq\ K(n)+K(x,p|\ \langle z,\langle n,K(n)\rangle\rangle\ ) -\mathcal{O}(1)$
\hspace*{\fill} (same expression, but unabbreviated)\\
$\geq\ K(n)+K(x,p|\ z') -\mathcal{O}(1)$
\hspace*{\fill} (via substitution $z':=\langle z,\langle n,K(n)\rangle\rangle$ )\\
$\geq\ K(n)+K(x|\ z')+K(p|\ z',K(z')) -\mathcal{O}(1)$
\hspace*{\fill} (via general. Chain Rule, Cor. \ref{chain rule corollary})\\
q.e.d.
\end{proof}
{\ }

Below, we will leverage the lemma above to prove an upper bound on $|\mathcal{V}_n(x|z)|$.
But first however, note that the number $n$ can be arbitrarily large and thus, 
in principle,
arbitrary pieces of information could be encoded within $n$.
Note furthermore that the information encoded within $n$ is also part of each $p \in \mathcal{V}_n(x|z)$, 
since any self-delimiting program $p$ carries the information about its own size $|p|=n$.
It would therefore be possible to unfairly manipulate the democratic outcome by purposefully selecting an $n$ that already contains information about a desired prediction. 
This problem can be resolved by disallowing $n$ from containing any information about either of the two candidate predictions $a$ and $b$, 
whenever calculating the probability ratio \mbox{(Eq. \ref{odds exp equation})} between these two candidates.
If $n$ does not contain any information about $x$ at all, $n$ will not be of much help when computing $x$ and therefore, $K(x|n)\approx K(x)$.
To avoid this vague sign ‘$\approx$', we can require that $n$ must fulfill the inequality $K(x|n)\geq K(x)$,
both, if $x=oa$ were true and also if $x=ob$ were true.
%these ‘and’ and ‘or’ are tricky in natural language, easy to be misunderstood here.
On the same grounds, we can furthermore require an extended version of this inequality to hold true, where, in addition to $n$, we also provide $K(n)$ as well as our independent variables $z$ as background information:
\begin{align}\label{neutrality eq}
    K(x|z,n,K(n))\geq K(x|z)
\end{align}
Any $n$ that fulfills this equation 
(both, if $x=oa$ were true and also if $x=ob$ were true)
may be called a \textit{neutral} size $n$ (with regards to $a$ vs $b$, and given $o$ and $z$).
\\

\begin{theorem}\label{Core}\textbf{Upper Bound on $|\mathcal{V}_n(x|z)|$ :}\\
For any neutral $n$, i.e.: For any $n$ fulfilling \mbox{Equation \ref{neutrality eq}}, the following inequality holds:

\[|\mathcal{V}_n(x|z)| \ \ \leq \ \ 2^{n -K(x|z) -K(n) + \mathcal{O}(1)} \]

\end{theorem}
\begin{proof}
$\ |\mathcal{V}_n(x|z)| \ = \
\sum_{p\in \mathcal{V}_n(x|z)} 1$
\hspace*{\fill} (trivially valid for any set)\\
$=\ 2^n\cdot\sum_{p\in \mathcal{V}_n(x|z)}\ 2^{-n}$
\hspace*{\fill} (trivially valid for any sum)\\
$=\ 2^n\cdot \sum_{p\in \mathcal{V}_n(x|z)} \ 
2^{-K(n)
-K(x| z')
-K(p| z',K(z'))
+\mathcal{O}(1)}$
\hspace*{\fill} (via Lemma \ref{nontrivial lemma})
\\
$=\ 2^{n
-K(x| z')
}
\cdot \sum_{p\in \mathcal{V}_n(x|z)} \ 
2^{-K(n)
-K(p| z',K(z'))
+\mathcal{O}(1)}$
\hspace*{\fill} ($K(x| z')$ is the same for all $p$)
\\
$=\ 2^{n
-K(x| z')-K(n)
}
\cdot \sum_{p\in \mathcal{V}_n(x|z)} \ 
2^{-K(p| z',K(z'))
+\mathcal{O}(1)}$
\hspace*{\fill} ($K(n)$ is the same for all $p$)
\\
$=\ 2^{n
-K(x| z')-K(n)
+\mathcal{O}(1)}
\cdot \sum_{p\in \mathcal{V}_n(x|z)} \ 
2^{-K(p| z',K(z'))}$
\hspace*{\fill} ($\mathcal{O}(1)$ is the same for all $p$)
\\
$\leq\ 2^{n
-K(x| z')-K(n)
+\mathcal{O}(1)}
\cdot 1$
\hspace*{\fill} (via Kraft Ineq., Thm. \ref{Kraft_conditional}, different variable names)
\\
$=\ 2^{n
-K(x| z')-K(n)
+\mathcal{O}(1)}$
\\
$\leq\ 2^{n
-K(x| z, n, K(n) )-K(n)
+\mathcal{O}(1)}$
\hspace*{\fill} (Since $z' := \langle z, \langle n,K(n)\rangle\rangle $ in Lemma \ref{nontrivial lemma})
\\
$\leq\ 2^{n
-K(x| z)\ -K(n)
+\mathcal{O}(1)}$
\hspace*{\fill} (via neutrality of size $n$, Equation \ref{neutrality eq})
\\
q.e.d.
\end{proof}
{\ }

Let us now combine both of our bounds on $|\mathcal{V}_n(x|z)|$
(\mbox{Theorems \ref{coreEasyDir} \& \ref{Core}}) to obtain the following theorem about $|\mathcal{V}_n(x|z)|$ with great ease:

\begin{theorem}\label{rangeCore}
\textbf{Equation for $|\mathcal{V}_n(x|z)|$ :}\\
For any $n$ which is not too small $($i.e., $n \geq K(x|z)+K(n)+\mathcal{O}(1))$
and \\ which is neutral $($i.e., $n$ fulfills \mbox{Equation \ref{neutrality eq}} $)$,
the following equation will hold:
\\
\[|\mathcal{V}_n(x|z)| \ \ = \ \ 2^{n -K(x|z) -K(n) \pm \mathcal{O}(1)} \]
\end{theorem}
\begin{proof}
This entire theorem follows trivially from combining \mbox{Theorem \ref{coreEasyDir}} with \mbox{Theorem \ref{Core}}.\\
q.e.d.
\end{proof}

\begin{corollary}\label{odds exp corollary}
\textbf{Equation for Probability Ratios :}\\
If, for both, whether $x=oa$ or $x=ob$, $n$ fulfills the initial requirements of \mbox{Theorem \ref{rangeCore}}, then, the following equation holds:
\[
\frac{|\mathcal{V}_n(oa|z)|}{|\mathcal{V}_n(ob|z)|}
\ =\ 2^{ K(ob|z)-K(oa|z)\pm\mathcal{O}(1)}
\]
\end{corollary}
\begin{proof}
This equation is easily obtained by, at first, applying \mbox{Theorem \ref{rangeCore}} to both, 
the numerator and the denominator:\[
\frac{|\mathcal{V}_n(oa|z)|}{|\mathcal{V}_n(ob|z)|}
\ =\ \frac{2^{n -K(oa|z) -K(n) \pm\mathcal{O}(1)}}
{2^{n -K(ob|z) -K(n) \pm\mathcal{O}(1)}}
\ =\ \frac{2^{-K(oa|z) \pm\mathcal{O}(1)}}
{2^{-K(ob|z) \pm\mathcal{O}(1)}}
\ =\ 2^{ K(ob|z)-K(oa|z)\pm\mathcal{O}(1)}\]
q.e.d.
\end{proof}
This corollary is a confirmation of the earlier \mbox{Equation \ref{odds exp equation}},
and therefore, 
validates the argument for Occam's razor described at the beginning of this subsection.
For a discussion of these deviations $\mathcal{O}(1)$ and more, see the following subsections.

\subsection{Discussion of Deviations and Incomputability}\label{deviations section}
In this subsection, 
we will resolve three objections to the practical applicability of Occam's razor.
Generally, one can expect various such objections to be raised by some of the scientists whose own models will be ruled out by the formulae above. 
Two of these objections, \textbf{(vii)} and \textbf{(viii)}, are based on the aforementioned deviations $\mathcal{O}(1)$.
Another related objection, \textbf{(ix)}, is based on the incomputability of Kolmogorov complexity.

Recall that the deviation $\mathcal{O}(1)$ in \mbox{Theorem \mbox{\ref{coreEasyDir}}} was bounded by a constant which is the size of the algorithm $r$ described in the proof.
Let us name it $C_\text{\ref{coreEasyDir}}$, 
after the numbering of the theorem, i.e.: 
$C_\text{\ref{coreEasyDir}} := |r|$ .
Analogously, the other constant, from \mbox{Theorem \ref{Core}}, 
we will call $C_\text{\ref{Core}}$.
Both of these constants can easily be carried forward into \mbox{Theorem \ref{rangeCore}} and \mbox{Corollary \ref{odds exp corollary}} to bound the deviations $\mathcal{O}(1)$.
The order of magnitudes of similar constants were previously found to be in the hundreds \cite{tromp2023functional}, 
and the same order of magnitude may thus be true for $C_\text{\ref{coreEasyDir}}$ and $C_\text{\ref{Core}}$.
Regardless of the exact size of these constants, 
the following objection may be raised:\vspace{1.5mm}
\\
\textbf{(vii)} Due to the deviation $\pm\mathcal{O}(1)$, 
\mbox{Corollary \ref{odds exp corollary}} will fail to inform us on whether the probability ratio will be in favor of $a$ or $b$, 
whenever the respective complexities are too close to each other, i.e., whenever the absolute difference $|K(ob|z)-K(oa|z)|$ is smaller than the constant $\pm(C_\text{\ref{coreEasyDir}}+C_\text{\ref{Core}})$, which bounds the deviation $\pm\mathcal{O}(1)$ from above and from below.\vspace{1.5mm}

This objection is largely irrelevant whenever the absolute difference between the complexities will be sufficiently much greater than this constant, such that the probability ratio will be guaranteed to be one-sided despite any possible deviations. 
However, the objection still raises the question of whether we can do better than this, i.e.: 
Can we increase the accuracy by getting rid of these deviations?
As we will show next, indeed, 
it does make sense to circumvent these deviations and constants, 
thereby making Occam's razor relevant also to smaller differences in model complexity.
Recall that for any given $n$, the probability $P(x|z)$ is proportional to the number $|\mathcal{V}_n(x|z)|$.
However in general, (frequentist-) probabilities are equal to the \textit{expected} rate of occurrence of some outcome,
which, in our case, is proportional to the expected value $\mathbb{E}( |\mathcal{V}_n(x|z)| )$. 
In order to obtain such an expected value,
one may introduce a probability distribution over the possible values of $|\mathcal{V}_n(x|z)|$,
restricted by the bounds provided in our previous theorems,
i.e., zero probability outside the bounds.
Let us write this bounded range of possible values as 
one interval, as follows:
\begin{align}\label{exponential interval}
    |\mathcal{V}_n(x|z)| \ \ \in \ \  
    [\ 2^{n -K(x|z) -K(n) -C_\text{\ref{coreEasyDir}}}\ ,\
    2^{n -K(x|z) -K(n) +C_\text{\ref{Core}}}\ ]
\end{align}
Let us rearrange this statement directly into the following, equivalent statement, where the entire left side is the deviation, which always falls within the interval that is determined by constants only:
\begin{align}\label{unrounded formula}
    \log_2|\mathcal{V}_n(x|z)|\ -n +K(x|z) +K(n) \ \ \in \ \ [-C_\text{\ref{coreEasyDir}} , +C_\text{\ref{Core}}]
\end{align}
If we round the logarithm up to the next integer, the statement still holds true:
\begin{align}\label{rounded formula}
    \lceil\log_2|\mathcal{V}_n(x|z)|\rceil\ -n +K(x|z) +K(n) \ \ \in \ \ [-C_\text{\ref{coreEasyDir}} , +C_\text{\ref{Core}}]
    \cap\mathbb{Z}
\end{align}
Note that the set $[-C_\text{\ref{coreEasyDir}} , +C_\text{\ref{Core}}]\cap\mathbb{Z}$ will be the same for any $x$.
Whatever one may assume the probability distribution (PD) of this rounded deviation over the set $[-C_\text{\ref{coreEasyDir}} , +C_\text{\ref{Core}}]\cap\mathbb{Z}$ to be, 
a priori, one should assume this PD over this set to remain approximately the same, regardless of whether $x=oa$ or $x=ob$,
since otherwise, this may be considered as an attempt to sneak in an unfair bias toward particular predictions,
especially if the respective complexities are close to each other (as in \mbox{Objection (vii)}).
Let $\varepsilon$ then be the \textit{expected} value of two to the power of the deviation.
Given that the unknown probability distribution remains the same for all $x$,
it follows that $\varepsilon$ will also be the same for all $x$. So, $\varepsilon$ will be the same, whether $x=oa$ or $x=ob$.
Therefore, when calculating the probability ratio for $a$ vs $b$, this deviation will cancel itself out as follows:
\begin{align}\label{odds accurate}
\frac{P(oa|z)}{P(ob|z)}
 := \frac{\mathbb{E}[|\mathcal{V}_n(oa|z)|]}{\mathbb{E}[|\mathcal{V}_n(ob|z)|]}
 = \frac{\varepsilon\cdot 2^{n-K(oa|z)-K(n)\pm 1}}
{\varepsilon\cdot 2^{n-K(ob|z)-K(n)}}
 = 2^{ K(ob|z)-K(oa|z)\pm 1}
\end{align}
The small deviation $\pm1$ just accounts for the rounding earlier, in \mbox{Equation \ref{rounded formula}},
but in \mbox{Equation \ref{odds accurate}},
note that, since $\varepsilon$ cancels itself out, 
the final term is now void of large deviations,
which means that this equation will inform us on whether the probability ratio will be in favor of $a$ or $b$, and by how much, even if the respective complexities are quite close to each other.
Ergo, \mbox{Objection (vii)} is hereby resolved.

However, concerning the validity of \mbox{Equation \ref{odds accurate}}, 
there is another objection that may be brought up against it. 
It goes as follows:\vspace{2mm}
\\
\textbf{(viii)}
Let us bring about a contradiction via the following example calculation.
For simplicity, assume the left side of \mbox{Equation \ref{unrounded formula}} (i.e. the deviation) to always be distributed such that it will be equal to $-100$ with a probability of $50\%$ and equal to $+100$ also with a probability of $50\%$.
Assume, furthermore $K(ob|z)-K(oa|z)=30$, and therefore, according to \mbox{Equation \ref{odds accurate}}, prediction $a$ is $2^{30}\approx$ 1 \textit{billion} times more likely to be correct than prediction $b$. 
However, since this difference, $30$, is smaller than the possible deviations ($\pm 100$) that may have opposite signs for $a$ and $b$,
it is possible, with a probability of $50\% \cdot 50\% = 25\%$, 
for the \mbox{ratio $\frac{|\mathcal{V}_n(oa|z)|}{|\mathcal{V}_n(ob|z)|}$} to be overturned by these opposite deviations, such that it would overwhelmingly favor $b$ over $a$,
i.e., $25\%$ chance of $\frac{|\mathcal{V}_n(oa|z)|}{|\mathcal{V}_n(ob|z)|} = 2^{30-100-100} = 2^{-170}$ being true.
This means that $b$ will be true with a probability of at least around $25\%$, 
which stands in direct contradiction to the probability of just about 1 \textit{billionth} that we just had calculated.
Therefore, in all cases where the difference between the respective complexities is too small, the high certainty, claimed in \mbox{Equation \ref{odds accurate}}, is unwarranted.\vspace{2mm}

Let us now resolve this paradoxical contradiction.
The resolution follows from the detection of a common fallacy called ‘false dichotomy'.
In particular, after having correctly deduced the fact that there is a case where the fraction could equal $2^{-170}$, 
it was implicitly assumed that, in this case, the probability of $b$ would be roughly equal to $1-2^{-170}\approx 1$.
However, such a step would only be warranted if $a$ and $b$ would be the only two possible predictions, but this is not so.
Therefore, in this case, the probability of $b$ could be orders of magnitude smaller,
and thus, after completing the calculation, the final result would not equal $25\%$.
Thus \mbox{Objection (viii)} is refuted and \mbox{Equation \ref{odds accurate}} remains valid.

When leveraging this equation to debunk existing models, we can expect yet another objection to be raised,
based on the incomputability of the Kolmogorov complexity,
which stems from the proven undecidability of the halting problem \cite{davis2013computability,bernays1936alonzo,turing1936computable}.
More precisely, Kolmogorov complexity $K(x)$ is said to be ‘upper semicomputable’ because some smaller programs will continue to run forever without returning an output where it is impossible to prove that such a program will \textit{not} suddenly return $x$ and halt \cite{davis2013computability,bernays1936alonzo,turing1936computable}, which would mean, per definition of $K$, that $K(x)$ may or may not be the size of that non-halting, smaller program, which cannot be determined.
For many $x$, this unpredictability makes it impossible to prove that $K(x)$ is not actually smaller than our best estimate, even if our estimate of $K(x)$ was already correct. 
This means that for many $x$, $K(x)$ can only be approximated from above but not from below, which is what ‘upper semicomputable’ means. 

Moreover, Chaitin's remarkable incompleteness theorem \cite{chaitin1974information,chaitin1977algorithmic,calude2013information} has shown that this incomputability of $K(x)$ applies to all possible $x$ with $K(x)$ greater than some constant, 
i.e.: In most cases, $K(x)$ could never be determined with full certainty.
Hence, the following objection to the practical applicability of \mbox{Equation \ref{odds accurate}} can be raised:\vspace{1.5mm}
\\
\textbf{(ix)} 
In practice, it will not be possible to utilize \mbox{Equation \ref{odds accurate}}, because, in most cases, 
it will be impossible to approximate Kolmogorov complexity from below, according to Chaitin's incompleteness theorem, and thus, it will be impossible to approximate the difference $K(ob|z) - K(oa|z)$ from above as well as from below.\vspace{1.5mm}

Objections of the same kind have previously been lodged by Sterkenburg \mbox{\cite[p. 651]{sterkenburg2019putnam}} and Neth \cite{neth2023dilemma}, without resolution.
Let us resolve this problem now.
Firstly, despite Chaitin's incompleteness theorem, it is a proven and well-known fact that most possible bit sequences are incompressible, i.e., $K(x)\geq|x|$ for most $x$ \mbox{\cite[Thm. 2.2.1]{li2008introduction}}\cite{kolmogorov1965three}.
This is because the possible bit sequences outnumber the smaller programs, such that most possible bit sequences will be left without a smaller program that will output them.
Similarly, if we already know a smaller program that will return some $x$, 
we should \textit{not} expect there to exist an even much smaller program that will also return $x$.
This is because the number of possible programs decreases exponentially with the length the programs \cite[Thm. 2.2.1]{li2008introduction}.
Therefore, the situation is not as dire as \mbox{Objection (ix)} may have made it seem.

Consider two candidate models, both consistent with past observations,
but one model predicts $a$ and the other model predicts $b$.
The former model may be older and overly complex and the latter model may be newer and simper. 
Proponents of the older model predicting $a$ may try to cast doubt on the new model by arguing that there may exist some significantly simpler version of their old model that also predicts $a$.
One would not be able to disprove this claim, 
as we know form Chaitin's incompleteness theorem.
However, as we just explained above, a program that would be significantly simpler reformulation of the old model is rather unlikely to exist \mbox{\cite[Thm. 2.2.1]{li2008introduction}}, and therefore, the burden of proof would rest with the proponents of the old model,
i.e.: They would actually have to bring forth a simper model that also predicts $a$, in order not to be displaced. 
Moreover, even simpler reformulations of the new model predicting $b$ could exist as well, 
and as such, the expected value of the difference $K(ob|z) - K(oa|z)$ equals the difference between the size of the known, new model and the size of the known, old model,
because any of the expected deviations will cancel each other out.
Therefore, the known sizes of the models already inform us about the relevant probability ratio required to displace the old model and hence, 
the incomputability will be irrelevant in practice,
which means that \mbox{Objection (ix)} is resolved now. 

In accordance with this discussion, since the exact Kolmogorov complexities are usually unknown, let us reformulate \mbox{Equation \ref{odds accurate}} as follows:
\begin{align}\label{odds computable}
\frac{P(oa|z)}{P(ob|z)}
\ =\ \frac{\mathbb{E}[|\mathcal{V}_n(oa|z)|]}
{\mathbb{E}[|\mathcal{V}_n(ob|z)|]}
\ =\ \frac{\mathbb{E}[2^{-K(oa|z)}]}
{\mathbb{E}[2^{-K(ob|z)}]}
\ \approx \ 2^{ K'(ob|z)-K'(oa|z)}
\end{align}
, where $K'(x|z)$ is the size of our smallest \textit{currently known} model that will return $x$ and halt, if given $z$.
$K'(x|z)$ is obviously always computable since we can formulate any theoretical model as a program based on the reference formalism (Section \ref{model complexity}) and then count the number of bits that this program consists of.

After now having established the validity and accuracy of a modernized Occam's razor under determinism, 
let us now raise the tenth objection:\vspace{1.5mm}
\\
\textbf{(x):} This whole proof of Occam's razor might be invalid because, 
besides all the deterministic models, 
there also exist many stochastic models, 
whose predictions were not taken into account in the democracy on which this proof was based.\vspace{1.5mm}

Most of the following subsection is dedicated to the resolution of this objection:

\subsection{Generalization to Stochastic Models}\label{stochastic section}
In this subsection, we show that the mathematical proof of Occam's razor will persist all the same when running a democracy of all possible stochastic models instead of only the deterministic models.
For the democracy of stochastic models, the resulting probabilities of observations will turn out to remain essentially the same as for the democracy of deterministic models.
A different question, which we will briefly touch on subsequently, is how to apply Occam's razor correctly when deciding between two stochastic models.

The loose term ‘stochastic models’ can mean one of two things:
Firstly, it may refer to nondeterministic/randomized algorithms \cite{motwani1996randomized} (e.g., Monte Carlo simulations \cite{harrison2010introduction}),
which are algorithms that rely on some source of randomness (or pseudorandomness) in order to produce varying results.
Secondly, it may also refer to formulae or algorithms, often called ‘probabilistic models', such as, for instance, models of quantum mechanics, whose output will be the probabilities of outcomes. 
Further examples would be statistical models used in various other experimental sciences.
Since the computation of such outcome probabilities is usually deterministic \cite{lin1988mathematics}, 
such probabilistic models may also be called ‘deterministic’ by some.
But for now, let us turn our attention to the less deterministic, randomized algorithms and let us call them ‘randomized models’ and we will get back to the ‘deterministic’ probabilistic models only thereafter.

Recall that in the previous subsections
we have represented each deterministic model as a self-delimiting program.
We could represent the randomized models in the same way but with the addition of a source of randomness, i.e., 
such programs should be provided with some randomly sampled bit sequence $\rho$.
Let $q$ be a program that represents a randomized model to which the noise $\rho$ should be provided.
One possible avenue for providing the noise $\rho$ to the program $q$ would be via another auxiliary input, 
akin to how the independent variables $z$ were provided in the previous subsections.
However, the subsequent argumentation will be shorter if $\rho$ is not provided alongside $z$, but if instead, $\rho$ is provided by appending it to $q$.
However, since a self-delimiting program cannot read its appendices, such as $\rho$, this means that we will cancel the requirement for program $q$ to be self-delimiting.
Recall that the original reason to require such self-delimitation was in order to avoid counting the entirely redundant, longer bit sequences,
that were mentioned in \mbox{Objection (vi)}.
In analogy, we require, for randomized models, that the concatenation $q$$\rho$ always be a self-delimiting program, rather than $q$ itself, 
in order to avoid \mbox{Objection (vi)} again.
The noise $\rho$ can be sampled via a sequence of fair coin flips, each of
which has a 50:50 chance to result in a zero or a one.
Due to the self-delimitation of $q$$\rho$, 
depending on the model $q$, the length of the noise $\rho$ may be long or short, as the model may contain instructions on how far the reference formalism should continue reading.
Furthermore, a model $q$ may also contain instructions on how far the reference formalism should continue reading based on properties of the noise $\rho$ itself.
This implies that the afore-mentioned process of sampling $\rho$ via fair coin-flips can lead to a probability distribution over possible lengths of $\rho$ itself.
One could therefore say that the probability distribution over these lengths $|\rho|$ is determined by the model $q$.
In order to prove Occam's razor again, we form a democracy of all possible randomized models $q$ of some great size $|q|=n$, each followed by some random noise. 
Then, let $n$ go towards infinity, to ensure maximal model complexity again. 
We can go one step further by demanding that the length of the noise $|\rho|$ go toward infinity as well, in order to ensure a maximal amount of randomization. 
This can be achieved by introducing another variable $l$, analogous to $n$, and by then simply rejecting any predictions that were output whenever length $|\rho|$ was not equal to $l$. 
%technically some voting opportunities is lost here: Semimeasures.
Then, also allow $l$ to go toward infinity, analogous to $n$, 
and see what the outcome of this fair, randomized democracy will be.
The probability with which some given model $q$ will output some $x$,
will therefore be proportional to the size of the set $\{\rho :\ |\rho|=l \ \land \ U(z,q\rho)=x\}$.
Taking into account the complete randomized democracy, 
the expected total number of different models $q$ that will output (or vote for) $x$ is going to be proportional to the size of the set
$\{q\rho :\ |q|=n \ \land \ |\rho|=l \ \land \ U(z,q\rho)=x\}$, which should be self-evident.
Given the program $q\rho$ as just a bit sequence, it is just a matter of interpretation, at which bit the model ends and the noise starts.
Since the reference formalism $U$ does not distinguish between the model and the noise, 
now use the substitution $p:=q\rho$ to simplify the aforementioned set as follows:
\[\{q\rho :\ |q|=n \ \land \ |\rho|=l \ \land \ U(z,q\rho)=x\}\ \ \
=\ \ \ \{p :\ |p|=n+l \ \land \ U(z,p)=x\}\]
Note that this last set can be represented by the notation $\mathcal{V}_{(n+l)}(x|z)$, via \mbox{Definition \ref{voteSetDef}}.
This means that we can apply the theorems of the previous subsections to this set as well.
Upon examination of \mbox{Corollary \ref{odds exp corollary}}, for the outcome of the democracy, whether the set is 
$\mathcal{V}_{n}(x|z)$ or $\mathcal{V}_{(n+l)}(x|z)$
will make no difference, as long as $n+l$ also follows our usual requirement of being neutral w.r.t. $a$ versus $b$.
Therefore the outcome with randomization will be identical to the outcome in the afore-discussed deterministic case, 
i.e.: It confirms Occam's razor.
Of course, if we allow for a larger range of lengths of noises $|\rho|$ to be represented simultaneously within the same democracy,
rather than just one length $l$, 
the final outcome will also still remain the same, as long as it is done in a reasonably neutral manner.
So, to summarize, despite us having provided maximal charitability toward arbitrarily high model complexities, as well as toward arbitrarily large amounts of randomization, 
the outcome will still end up agreeing with Occam's razor all the same.

Within the field of algorithmic information theory, various similar results on the correspondence between stochastic programs and deterministic programs and probability distributions have long been known \cite[Chap.5]{devine2020algorithmic}\cite[Thm.4.3.3]{li2008introduction}\cite[Eq.5.5]{hutter2005universal}\cite{mueller2020law}. 
Based on such results, it can be shown that Occam's razor will continue to hold true if we use consider all possible probabilistic models whose output will be numbers that represent probabilities of outcomes.
This is because, loosely speaking, randomized models can be converted into other probabilistic models and vice versa, via constant-sized algorithms,
but our detailed discussion of the randomized models above should already suffice to illustrate the point that stochasticity will not affect the veracity of Occam's razor at all. 
Therefore, \mbox{Objection (x)} is resolved.

Another question is how to apply Occam's razor correctly to decide between two stochastic models,
as it could be wrong to select a stochastic model solely based on its simplicity, since the simpler model might, in turn, have permitted greater deviations from past observations.
The very same question can be raised in the case of 
\textit{imprecise} deterministic models, i.e., 
deterministic models that do not reproduce past observations $o$ precisely, 
but may nevertheless be useful due to them matching past observations $o$ approximately.
There exist multiple approaches for resolving such issues and multiple different terminologies exist \cite{tian2022comprehensive,li2003sharpening}\cite[Chap. 5]{devine2020algorithmic}. 
Here, we will only introduce the common idea of these approaches,
which is to add up two quantities, both of which should be minimized together.
This is analogous to our sum $n+l$ above.
the first quantity is the model complexity and the second quantity is the ‘empirical loss’ $\mathcal{L}_\text{emp}$, also known as ‘prediction error'.
$\mathcal{L}_\text{emp}$ quantifies how strongly the model's imprecise output $o'$ deviated from the true past observations $o$.
Most generally, one could express $\mathcal{L}_\text{emp}$ as the estimated additional amount of information $K'(o|o')$ that would be required to fully correct the model's imprecise output from $o'$ to $o$.
Many practical methods exist to compute an empirical loss, which, in practice, is often as simplistic as calculating the Hamming distance between $o'$ and $o$. 
In the different case of a model whose outputs are probabilities, 
one can obtain $\mathcal{L}_\text{emp}$ by taking the logarithm of the probability that the model assigned to $o$,
as is common in machine learning\footnote{
Such logarithms commonly appear within the ‘cross-entropy loss’ \cite{de2005tutorial}, which is a standard practice in modern machine learning.}.
Such log probabilities are also called the ‘surprisal'.
If the model complexity is added to $\mathcal{L}_\text{emp}$, the result is commonly called the ‘regularized loss’ $\mathcal{L}_\text{reg}$.
So, the regularized loss $\mathcal{L}_\text{reg}$ also includes the model complexity. 
Now, when applying Occam's razor to decide correctly between stochastic models, 
one ought to choose the stochastic model with the smallest \textit{regularized} loss $\mathcal{L}_\text{reg}$, rather than just the simplest stochastic model.
The same holds true outside of machine learning and also applies to any scientific models, including theoretical physics formulae.

\subsection{Discussion of Initial Assumptions}\label{assumptions}
Here, we will gather and discuss all of the initial assumptions, premises, postulates, presumptions, or principles,
on which we have built our mathematical proof of Occam's razor earlier in this paper,
and thereafter, we will briefly review previous authors' alternative initial assumptions, 
which may have been more restrictive, too strong, or superfluous.

\subsubsection{Initial Assumptions of the Proof}\label{our own assumptions}
We introduced three initial assumptions within the first paragraph of \mbox{Subsection \ref{models_as_algorithms}}, which we will paraphrase here:\vspace{1.5mm}
\\
\textbf{1.} Reproducibility: Scientific models have to meet the standard of reproducibility \cite{gundersen2021fundamental,national2019reproducibility}.\vspace{1.5mm}
\\
\textbf{2.} The Church-Turing thesis 
\cite[Thesis 2.3]{hutter2005universal}
\cite{sep_church_turing,turing1936computable,bernays1936alonzo}.\vspace{1.5mm}
\\
\textbf{3.} Full Generality: No further restrictions on what constitutes a valid scientific model.\vspace{1.5mm}
\\
Among computer scientists, the Church-Turing thesis is a settled consensus, so we will not discuss it further at this point. 
Our generality assumption can be regarded as a variant of Epicurus' principle of multiple explanations \cite{EpicurusMulti84}, 
which states that all possible explanations/models are to be retained, as long as they are consistent with observations at least.
In this context, whether implicitly or explicitly, these three assumptions were being made repeatedly, ranging back to Solomonoff \cite{solomonoff1964formal1,solomonoff1964formal2,rathmanner2011philosophical}.
Our fourth initial assumption can be found in our \mbox{Subsubsection \ref{circular resolution}}:\vspace{1.5mm}
\\
\textbf{4.} Untyped lambda calculus \cite{barendregt1984lambda,bernays1936alonzo} is extremely simple, according to common sense.\vspace{1.5mm}
\\
Note that technically, we did not directly assume the untyped lambda calculus (ULC) to be extremely simple,
but instead, we made the even more modest assumption, which is that according to common sense, the ULC is extremely simple.
We think that any philosopher will be able to agree with this weak assumption.
Later within the same subsubsection, 
we justified the demand for the simplicity of the reference formalism,
by positing the need for some reliability when measuring of complexity.
This is because allowing for complicated reference formalisms would open the door toward sneaking unfair biases into the measure of complexity, which would enable false claims of certain complex models being simple.
Therefore, the initial assumption on which our argument was based may be called neutrality, fairness, objectivity, impartiality, or something to that effect.
Similar assumptions, based fairness, then appeared at least four more times, throughout \mbox{Section \ref{main section}}, which we will review now.

In \mbox{Subsection \ref{Core Section}},
we assumed that a democracy of models should be fair in the sense that all votes will carry equal weight, ‘one model one vote!'.
In the deterministic case, no sensible alternative is known, and as we have just shown in \mbox{Subsection \ref{stochastic section}}, 
the stochastic approaches always end up agreeing with the deterministic approach. 
Next, we have argued that $n$ should fulfill \mbox{Equation \ref{neutrality eq}}, 
based again on the assumption that one is supposed to ensure neutrality or fairness. 
In principle, it would alternatively also have been possible to formulate the proof of Occam's razor without requiring $n$ to fulfill \mbox{Equation \ref{neutrality eq}},
but this would just have made the proof more complicated.
In \mbox{Subsection \ref{deviations section}},
before \mbox{Equation \ref{odds accurate}},
we argued yet again based on ensuring fairness, 
that between two different predictions ($a$ vs $b$),
the two probability distributions we assume over the same interval should be the identical. 
Finally, in \mbox{Subsection \ref{stochastic section}}, we made the optional assumption of the random noise being generated by fair coin flips.
Now, let us reduce all of these five assumptions into the following single, initial assumption:\vspace{1.5mm}
\\
\textbf{5.} Impartiality/neutrality/fairness/objectivity is to be pursued, in science/epistemology.\vspace{1.5mm}
\\ 
This is also related to the principle of indifference \cite{keynes1921chapter},
which Laplace considered to be obvious and which previous authors have also used to argue for Occam's razor \cite{rathmanner2011philosophical}.

These five initial assumptions that we have stated above are all of the initial assumptions that we have identified as the foundation for our general mathematical proof of Occam's razor. 
Consequently, anyone who agrees with just these five initial assumptions is invited to agree with our results and conclusions.

\subsubsection{Alternative Initial Assumptions}\label{superfluous}
In this subsubsection, we shall briefly review previous authors' alternative initial assumptions, 
which may have been more restrictive, too strong, or superfluous.

An early paper that argued for the validity of Occam's razor based on Kolmogorov complexity
started from the assumption that a hypothetical ‘Great Programmer’ had once started to run a simple program on his big computer \cite{schmidhuber1997computer} 
that generated all possible computable universes \cite{schmidhuber2012fastest}, 
i.e., a most general, simple multiverse, which we inhabit.
However, since our five initial assumptions above (\ref{our own assumptions}) should already have sufficed to prove Occam's razor, 
the ‘Great Programmer’ is an unnecessary initial assumption.
The same paper also assumed our universe to be computable.
The ‘Great Programmer’ assumption was not adopted by Hutter, 
but the assumption that our universe is computable did survive into his argumentation for Occam's razor. 
He states: “Assume we live in the universal multiverse that consists of all computable universes” \cite{hutter2010complete_occamProof}.
Even this assumption, however, still seems to be too strong of an initial assumption, 
thus potentially reducing the credibility of the argumentation.
Regardless of how one may interpret the meaning of this assumption, 
is it really necessary to make such an initial assumption about the universe?
Not quite, because instead, 
it should suffice that the abstract/theoretical modeling capabilities of reproducible science are ultimately restricted to computable models only, 
and therefore, the contemplation of alternatives is futile,
and thus, the only option left is to follow the democracy of models, 
which then confirms Occam's razor.
For a discussion of models that are not just abstract/theoretical, 
(i.e. physical scale models made from literal matter) see our \mbox{Subsubsection \ref{scale_models}}.

Another initial assumption made by Hutter is the ‘short compiler’ assumption\\
\mbox{\cite[Assumption 2.5]{hutter2005universal}},
which essentially states that all \textit{natural}, computationally universal formalisms will always be able to emulate each other via reasonably small programs.
This assumption leaves room for multiple interpretations.
Müller later started to reason from similar assumptions, 
only to then discover that these assumptions were false \cite{muller2010stationary}.
Nevertheless, it should obviously be true that all formalisms that would be well-suited as \textit{reference} formalisms should be able to emulate each other via reasonably small programs. 
However, rather than this being an initial assumption, 
such a statement already seems to follow from the lines of reasoning in \mbox{Subsubsections \ref{circular resolution} and \ref{lambda}}.
Moreover, barely any universal formalisms as simple as the ULC even exist, 
but for the sake of the argument, 
let us make the unrealistic, hypothetical assumption that there exists an unknown, well-hidden mystery reference formalism as simple as the ULC, 
but which the ULC could only emulate via an excessively large program.
According to the invariance theorem \mbox{(Subsubsection \ref{lambda}),} 
the deviations between the two alternative measures of complexity would be bounded by a constant the size of this large emulator program.
Due to this constant bound, this deviation could be treated as just another addition to the deviations that we already discussed in \mbox{Subsection \ref{deviations section}}, which should be ignored in practice,
as we have argued when resolving Objections (vii) and (viii).
Therefore, no version of the short compiler assumption seems to be necessary.
But to return from this hypothetical scenario back to reality, 
some other simple, known universal formalisms have actually been investigated by Tromp \cite{tromp2023functional}, 
who found them to be less well-suited as a reference formalism than the ULC/BLC.
But let us now move on to discuss the next alternative initial assumption.

The interdisciplinary paper by \mbox{Dingle et al.} \cite{dingle2018input} 
also employed Kolmogorov complexity in a similar fashion to our paper in order to explain why various complex systems, 
ranging from biochemistry to financial markets, can be understood via simpler models.
Thereby, a restrictive assumption was made:
The output of any model will contain much more information than the model itself \cite[Suppl. 3]{dingle2018input}.
The reason that they introduced this initial assumption is essentially the same reason why we required \mbox{Equation \ref{neutrality eq}} to be fulfilled. 
However, note that our \mbox{Equation \ref{neutrality eq}} is much less restrictive than their assumption. 
\mbox{Equation \ref{neutrality eq}} allows models to contain \textit{arbitrarily} larger amounts of information than their output.
This freedom was instrumental for our argument, 
whose starting point was maximally charitable toward \textit{arbitrarily} complex models,
in order to maximize the credibility of the argument for Occam's razor, 
as the proof works in spite of all this charitability.

\section{Fundamental Physics and Our Universe}\label{Physics section}
The main purpose of this section is to introduce important improvements to the methodology of the theoretical physicists who work on the foundations of physics.
We first address two more objections to Occam's razor in \mbox{Subsections \ref{universe complexity}} and \ref{hierarchy}, related to physics.
In \mbox{Subsection \ref{stagnation}}, 
we then shed light on the root cause of the present stagnation of the field of fundamental theoretical physics, 
and in \mbox{Subsection \ref{alleviate}}, 
we go on to explain how and why this stagnation could realistically be alleviated via the adoption of a novel quality standard into the theoretical research practice, 
namely, metamathematical regularization.

\subsection{On the Complexity of our Universe}\label{universe complexity}
Let us start with the following straightforward objection to Occam's razor:\vspace{1.5mm}
\\
\textbf{(xi):} 
The formulae above show increasingly complex outcomes to be exponentially less likely.
Therefore, if Occam's razor or these formulae were true, a priori, we should expect to find ourselves in the simplest possible universe that allows for observers to exist, 
or at least, we should expect all of our observations to be extremely simple,
which directly contradicts reality, since we can observe the world around us to be highly complex.\vspace{1.6mm}

We should firstly point out that there is more than one paradox at play here.
Firstly, the Kolmogorov complexity ($K$) of our universe could much smaller than it appears, as was suggested by Tegmark \cite{tegmark1996does}. 
In fact, the $K$ of our universe could be equal to merely the amount of information present in the fundamental laws of physics, plus the amount of information present in the initial conditions of our universe.
This is because, given the initial conditions and the laws of physics, one could compute all of the events that have ever occurred in our spacetime, including all of the complex details that surround us. 
Both, the true laws of physics as well as the initial conditions are \textit{not} known to be highly complex and are both likely orders of magnitude simpler than our environment appears to be.

One may object yet again by pointing out that quantum mechanics is inherently nondeterministic and therefore, 
the initial conditions combined with the laws will still be insufficient to accurately compute all of the subsequent events.
However, there exist valid, deterministic interpretations of quantum mechanics, such as for instance, Everett's many-worlds interpretation\footnote{
Everett’s many worlds interpretation is not to be dismissed via Occam's razor (at present), if interpreted correctly, since the small size of the source code is relevant, while the vastness of the computational resources required to run many worlds is irrelevant. 
%Old text of this footnote: Schmidhuber's speed prior \cite{schmidhuber2000algorithmic,schmidhuber2002speed,schmidhuber1997computer} rejects Everett’s many worlds interpretation (MWI). However, this rejection was obsoleted by Hutter's CToE selection principle \cite{hutter2013subjective}. Therefore, the dismissal of the MWI is not warranted: Physicists should not take the MWI too lightly.
}
\cite{dewitt2025many}. 
These facts already suffice to show \mbox{Objection (xi)} to have been based on unwarranted claims of high complexity.
Such facts were also pointed out by Tegmark \cite{tegmark1996does} and Schmidhuber\footnote{
We do not rely on the strong initial assumptions made in \cite{schmidhuber1997computer}. See our Subsection \ref{assumptions}.
} \cite{schmidhuber1997computer,schmidhuber2000algorithmic},
who both furthermore noted that, in principle, various kinds of multiverses could have an even lower $K$ than a universe that is a constituent of such a multiverse \cite{tegmark1998theory}.
However, such multiverse theories will often be lacking in predictiveness, as they allow for a broader range of possibilities. 
This problem was resolved by Hutter \cite{hutter2013subjective}, 
who pointed out that \textit{complete} multiverse models should not only model the physics, but should additionally also model where and how our observations are obtained in the multiverse. 
So ultimately, such a complete model, including this ‘observer localization’ model, is what Occam's razor should be applied to, rather than just the physics model.

This brings us to our next point, which is that not only the $K$ of our world, 
but also the $K$ of our observations thereof could be much smaller than they appear. 
Take for instance some observation that consists of gigabytes of video recordings.
In principle, such observation could be generated by an algorithm consisting of the small algorithm that computes our entire universe's history at a microscopic resolution,
combined with another algorithm that will search this computed spacetime for the device that holds/performs the recording,
based on some search criteria that could consist of much less data than the video itself.
Once the algorithm finds such a device, it extracts the recording from the device, outputs the video, and halts.
So the true $K$ of the video could be orders of magnitude smaller than gigabytes,
as had long been understood by Hutter \cite{hutter2010complete_occamProof,hutter2013subjective}.
Considering these facts, \mbox{Objection (xi)} is thus invalidated more thoroughly.

Now, we could go one step further in this direction:
As \mbox{Objection (xi)} suggested, due to our previously proven formulae, 
we may assume that any epistemically rational observer should a priori expect the $K$ of all their direct observations to be small as possible,
and thus, should expect the fundamental laws of physics plus the initial conditions to be as algorithmically simple as possible. 
However, at this point, another paradox comes into play. 
Despite our formulae and Occam's razor being true, 
the expected value of the $K$ of our observations should, in fact, be infinite.
This is because the probability distribution over observations $P(x|z)$, appearing in \mbox{Equation \ref{odds accurate}},
cannot deviate by more than a multiplicative constant from Solomonoff's universal prior $m(x|z)$,
whose Shannon entropy is known to be infinite \cite[Exercise 4.3.4]{li2008introduction},
which proves the infinity of the expected $K$.
To save Occam's razor, one then may try explain away this infinity by pointing out that we may live in a nondeterministic universe with simple laws and that the infinite expected $K$ just represents the infinite expected amount of random information contained in the truly random noise of simple stochastic models.
However, While this is indeed a valid explanation of the infinite expected $K$,
this explanation does not rule out that the expectation value of the complexity of the laws of physics themselves (excluding the noise) could also be infinite.
This means that the laws of physics being complicated could be equally plausible to the laws of physics being simple.
This appears to be unexplored territory, but be that as it may,
this paradox is resolved if one recognizes that Occam's razor and our formulae are weaker, humbler statements, in the following sense.
Rather than prescribing some kind of simplicity to our world, 
the formulae merely state that, when choosing between two predictions, the one supported by the simpler model is much more likely to be correct,
but note that there could be such a large number of predictions lacking simple models, that their sheer number will counteract their lower individual probabilities, such that the sum of all their probabilities will make it at least plausible for the laws of physics to be quite complicated.
However, also note that this large set of predictions, lacking simple models, is not useful in practice, 
since its many different, unlikely predictions all contradict each other.
But on the other hand, 
if we only consider some features\footnote{
While an informal notion of ‘feature’ should suffice for the comprehension of our text, 
there also exists a formal notion of ‘feature’ \cite{franz2021theory}.
For a formal reformulation into a democracy, 
introduce an algorithm that will recover any feature from $x$, given its residual, whenever the feature is indeed a property of $x$.
Each voter consists of said algorithm, a residual, and  $x$.}
of each prediction, rather than entire predictions, large fractions of these predictions can agree on some features.
However, such a situation can be reformulated back into the democracy (Sec. \ref{Core Section} \& \ref{stochastic section}) of many complex models, and thus, 
for whatever feature the many predictions had agreed on, 
this feature will already have been predicted by a simple model, i.e.: 
Occam's razor obsoletes such attempts to make predictions based on the large set of predictions that lack simple models.
This means that, despite the plausibility of there being arbitrarily complicated laws of physics,
one must nevertheless utilize a modernized Occam's razor in order to drastically increase one's probability of obtaining correct predictions and alternative approaches are futile. 

Concerning the big bang, 
since thermodynamic entropy can also be considered to be a measure of information, 
one may ask whether the ‘a priori plausible’ algorithmic simplicity of our universe would already suffice to directly imply that the thermodynamic entropy of our early universe has had to have been extremely low.
The answer is that the relation between $K$ and thermodynamic entropy is nontrivial \cite{zurek1989algorithmic,devine2020algorithmic,ebtekar2025foundations} and outside the scope of our paper.
For now, at the very least, the a priori plausibility of a low $K$ makes the low entropy less surprising,
such that the observed, low, initial thermodynamic entropy should not mystify physicists quite as deeply anymore.
The same is analogously true for the relatively small physical size of the early universe.

\subsection{Hierarchical Decomposition of Research Problems}\label{hierarchy}
Even after having accepted the general validity of a modernized Occam's razor, 
physicists may claim that this principle might currently not be useful to them, for the following reason;
this is the 12th and final objection discussed in this paper:\vspace{1.6mm}
\\
\textbf{(xii):} 
Granted, some time in the future, Occam's razor may become relevant for deciding between multiple theories contending to be recognized as the correct so-called ‘theory of everything’ (ToE) in physics.
However, at present, a completed, viable ToE-candidate does not exist.
The quest for the ToE can be subdivided into easier subproblems, each of which consists in explaining only some incomplete subset of all physical phenomena, e.g., only explaining gravity. 
Now, if we were to apply Occam's razor to each of these subproblems separately, to find simple models for each individually,
why should this strategy ultimately help us discover the simplest ToE?
This strategy, based on multiple ‘incomplete’ models, might be \textit{misaligned} with the ultimate goal of finding the simplest, complete model, that solves all of these subproblems simultaneously, i.e., the ToE-model.
Therefore, Occam's razor may be largely irrelevant for advancing the foundations of physics at present.\vspace{1.7mm}

Let us now counter this objection by showing that the individual application of Occam's razor to each subproblem separately is, per se, already the correct strategy\footnote{However, physicists would have to execute this strategy more thoroughly and more rigorously than they have been doing so far.
See {Subsection \ref{alleviate}.}}.
We shall keep this counter argument brief and informal, but such that anyone who studied our entire paper shall be able to understand.
Each of the subproblems could be represented as a select sequence of some observed outcomes $o$, optionally, in combination with corresponding independent variables $z$.
For each subproblem, this data would have been selected to contain only the information relevant to the subproblem.
This data may stem from real experiments,
but it is also admissible for such data to have been synthesized to realistically mimic some known properties of our universe. 
We know that, among the many models that are solutions to such a subproblem, 
there also exist some models that contain the correct ToE-model.
However, these models, containing the correct ToE-model, are too difficult to discover and are themselves often not the simplest solution to the subproblem, 
especially if the subproblem concerns only a small fraction of the known aspects of physics.
From the formulae in \mbox{Subsection \ref{deviations section}}, 
we have learned that predictions that are supported by simpler models, should be expected to agree with an exponentially larger number of complex models.
Therefore, the pursuit of ever simpler models that solve a subproblem
will also lead to the exponential increase in the expected percentage of models that contain the correct ToE-model, among models that will agree with our simplest known model on various predictions.
%The proof of the lower bound in \mbox{Theorem \ref{coreEasyDir}} is based on models that could contain the ToE-model within $g$ in some form, but $g$ has no influence on the outcome of the computation. 
%However, the simplest solution to the subproblem could itself already be the ToE also, so this is no problem.
From this exponential increase also follows an exponential increase in the probability that our simplest known solution to the subproblem will show some resemblance to the correct ToE-model in certain aspects, such that it may serve us as a so-called ‘toy model’ \cite{reutlinger2018understanding},
whose investigation would allow for advances in the comprehension\footnote{Thus, Chaitin's aphorism “Compression is Comprehension” \cite{zenil2020compression,chaitin2006limits}.}
of the true nature\footnote{
Since the field of mathematics often revolves around the study of \textit{simple} abstract objects, 
this explanation of ours should suffice to enable the reader to finally make rational sense of “The Unreasonable Effectiveness of Mathematics in the Natural Sciences” \cite{wigner1990unreasonable}.
} of our universe.
So, we have now shown this strategy to lead to an increase in the probability of finding the correct ToE, i.e.: 
The strategy is aligned with the ultimate goal.
In this sense, \mbox{Objection (xii)} is now refuted.

For a problem as difficult as finding the ToE, 
one should obviously pursue a hierarchical strategy:
Tackle the easiest subproblems first in order to then use them as stepping stones that will ease the difficulty of finding the simplest models for increasingly harder subproblems, involving increasingly more laws of physics. 
However, there is no strict guarantee that a solution to a subproblem will serve as a useful stepping stone. 
Yet, according to the ‘optimal ordered problem solver’ strategy \cite{schmidhuber2004optimal}, 
it would be rational to spend only about half of ones time trying to come up with a completely new model and one should spend the other half of ones time on trying to come up with a model that incorporates aspects of the simplest models that have already solved related, easier subproblems.

Lowest in such a hierarchy would sit the smallest subproblems, each of which may concern the modeling of just a single symmetry, such as Gauge symmetry or Lorentz symmetry.
Higher up in the hierarchy would sit various larger subproblems, concerning various combinations of such symmetries, 
and at the top of this hierarchy would obviously sit the main problem of finding the correct ToE.
Obviously, research work on (sub-)problems that sit high in this hierarchy (e.g., some model of quantum gravity)
is supposed to try to incorporate the simplest models that already solved subproblems that sit lower in the hierarchy, 
e.g., the simplest model from which some symmetry group emerges.
One may assume this to already be the standard practice, but this is not so.
In reality, many physicists were unaware of the true power of Occam's razor, i.e., the exponential relations shown in \mbox{Equation \ref{odds computable}}.
Furthermore, physicists have usually never bothered to actually calculate their model complexities,
thus rendering the discussed strategy so inefficient that it caused the infamous multi-decade-stagnation of this field.
We will better explain our diagnosis in the following subsection.

\subsection{Stagnation of Fundamental Theoretical Physics}\label{stagnation}

Concerning the half-century-long stagnation \cite{hossenfelder2018lost,smolin2007trouble,woit2006not,unzicker2013bankrupting} of the research field of fundamental theoretical physics,
we posit that the amendable root cause thereof is the theoretical physicists' omission of calculations of the complete algorithmic information contents of their theoretical models, 
i.e., omissions of model complexity calculations.
If this defect were to be rectified, it could cause the rate of progress to be increased, 
possibly by orders of magnitude.
Here, we will first explain why the old research methodologies are insufficient.
We then explain how this state of affairs should be rectified in \mbox{Subsection \ref{alleviate}}.

Within any sophisticated research field, 
the intelligent allocation of various resources, such as time and money, 
depends on a great number of decisions at multiple scales.
These scales are called ‘levels of analysis’ by economists \cite{dopfer2004micro}.
At the smallest scale, the micro level,
there are the time resources of the individual researcher, 
who repeatedly decides on questions such as,
which of their own ideas to spend their time thinking and writing about,
and which of their peer's papers to spend their time reading and contemplating.
At an intermediate scale, the meso level,
there are the resources of small organizations,
who repeatedly decide on questions such as, 
which researchers to hire and fire, and which researchers to award prizes to.
At the largest scale, the macro level, there are the resources of nations, who decide on questions such as,
whether to fund the LHC, and whether to fund entire subfields, e.g., string theory. 

Note that for each of these questions, at all these levels of scale, the optimal decision making will depend on estimating how promising which research directions are. 
Ideally, the amount of resources allocated to any research direction should be proportional to its probability of leading to successful predictions \mbox{\cite['bias-optimality']{schmidhuber2004optimal}}\cite{levin1973universal,levin1984randomness}.
Since we already know the ratios between such probabilities to scale exponentially with differences in algorithmic complexity, 
this is an extremely powerful guiding principle for the intelligent allocation of resources.

Obviously, to enable such optimal resource allocation policies, physicists would first have to be able to correctly quantify their model complexities. 
However, unfortunately, the physicists have still been using ancient to centuries-old methodology for quantifying model complexity, 
leading to a rather poor tradeoff between accuracy and efficiency.
Because these old methods are nevertheless loosely correlated with the true model complexities,
the field was nevertheless able to progress before the onset of the stagnation.
One of these old methods is to simply count the number of postulates one which a model or theory is based,
but the number of postulates will an be inaccurate measure of information because one postulate can itself contain arbitrarily much more or less information than another postulate.

Another old method is to rely on the superficial simplicity of mathematical formulae, 
without taking into account all the information hidden in the mathematical definitions and axioms that were required to give meaning to the simple formulae.
A related old method is to pursue aesthetically pleasing, elegant mathematical formulae \cite{hossenfelder2018lost,smolin2007trouble}. 
The reason that such beauty was nevertheless correlated with simplicity, thus sometimes yielding success to physicists, 
is that the perception of beauty can be explained in terms of data compression by (natural) neural networks \cite{schmidhuber2008driven}, 
which is obviously somewhat related to algorithmic simplicity, yet still inaccurate for estimating model complexity.
The pursuit of such mathematical beauty has been deemed insufficient by Hossenfelder \cite{hossenfelder2018lost} and Smolin \cite{smolin2007trouble}.

There also exists an old method, that may be capable of higher accuracy, but is, in turn, much more time-consuming, and ultimately, too inefficient.
This method is for a smart individual to spend lots of time carefully reading and contemplating each idea/paper/project, to achieve such a deep, intuitive understanding of each, 
that they may eventually be able to instinctually, arrive at the correct assessments, based on their brain's inherent ‘Bayesian Occam's razor’ \cite{tenenbaum2011grow,chater2003simplicity}.
They will then be able to recommend the papers with the better solutions to the rest of the community.
However, the time of any individual will usually be insufficient to carefully study all of the promising papers on one topic.
Combine this with the fact that, even if it were hypothetically possible for an individual to carefully compare all of the many relevant papers to each other, such paper recommendations would still not constitute reliable evidence to other physicists, since they cannot verify the recommendation's impartiality nor can they verify how much thought was behind the recommendation. 
All of this means that novel theoretical insights will usually not be taken very seriously and usually, rightfully so.
This lack of well-directed attention is further exacerbated by the wide-spread unawareness of the true power of a modernized Occam's razor, i.e., the exponential relations shown in \mbox{Equation \ref{odds computable}}.

Consider the case of a large number of different, alternative solutions to the same problem, of explaining the same physical phenomenon. 
Each solution may be described in a different, technically sophisticated paper.
Ideally, the most promising solution would rise to the top, garnering the attention of the research community, who would then try to incorporate these insights into their other models, as discussed in the previous subsection. 
However, considering the aforedescribed, multiple inaccurate/inefficient/untrustworthy methods for determining model complexity presently in use, 
in combination with the interrelated, wide-spread unawareness of the aforementioned true power of a modernized Occam's razor, 
one cannot expect the best solutions to be recognized as such, by the relevant communities, within any reasonable amount of time.
The most valuable theoretical insights thus fail to be incorporated into other models, which is what severely hampers the progress. 
This exactly is the stagnation. 
Upon further contemplation, one can see that the same shortcomings apply to all of the aforementioned levels of analysis, 
from the numerous small ideas of individuals, all the way up to the impactful decisions of governments, and to all of the research directions simultaneously, 
thus causing the unfortunate stagnation of the entire field of fundamental theoretical physics.
So, we already see the root cause of the stagnation, 
without needing to consider the even more unpleasant fact that different factions may be incentivized to hide the weaknesses of their respective approaches when competing for funding, thereby further obfuscating excessive model complexities, as is enabled by the aforementioned old methods.

For any phenomenon in fundamental physics, do we currently know what the objectively simplest, published model is? 
The honest answer here would be ‘No', because there do not exist any official, accurate rankings of models based on their complexity, at present. 
But as shown in this paper, this field is supposed to be actively searching for the simplest models, as this could easily increase the chances of forthcoming success by orders of magnitude, 
as shown by the aforementioned exponentials.
So then why should one expect this field to ever progress, 
if it currently fails to identify its simplest models?
The answer may be: Because there used to be significant progress sometime in the past.
However, by now, this time lies about half a century in the past.
Ergo, one may have to estimate this stagnation to persist for one more half of a century into the future\footnote{
Via the naive application of Gott's temporal Copernican principle \cite{gott1993implications}.
}, ultimately totaling an entire century---that is, 
unless a metascientific paradigm shift is imminent. 
In what follows, we will explain how such a paradigm shift could realistically be achieved at present.

\subsection{Methodology for the Alleviation of the Stagnation}\label{alleviate}
In this subsection, we explain how the stagnation of the field of fundamental theoretical physics could possibly be alleviated much sooner than would otherwise have been the case.
As we will show here, the main methodological upgrade that the physicists would need to introduce now, 
is for them to start actually calculating all of their model complexities and to report these numbers clearly visible in their papers, 
in an official ‘measurement unit’ of information, 
such as bits or nats \cite{IEC80000132008,cover1999elements}.
These numbers would serve as a basis for an objective measure of theoretical significance, that would enable the quick and easy comparison between the significance of different papers.
Such a greatly increased efficiency would allow for official rankings\footnote{
Similar to how AI algorithms were long being ranked, thereby fostering fast progress.
} of the many theoretical solutions.

The reporting of these numbers would then be analogous to the long-established quality standard of reporting statistical significances in experimental science.
Without such a quality standard, experimental science would probably also have stagnated over the past hundred years\footnote{
One hundred years ago, statistical significance testing was successfully introduced to researchers of many fields, by Fisher \cite{fisher1925statistical}.
It was originally invented earlier by Pearson \cite{pearson1900x}.
}, for analogous reasons to the ones explained in the previous subsection.
This is why, over time it has become commonplace for experimental scientific journals to require statistical significance reporting (or similar), 
thereby enabling the great series of successes across many fields of science, which is also apparent in present-day fundamental \textit{experimental} physics. 
Analogously, in order for such success to be achieved in fundamental \textit{theoretical} physics, 
let us now explain how to correctly calculate any theoretical physicist's model's complexity,
and why it would be practically feasible to do so.

While being practically feasible, the new calculation method must avoid all of the shortcomings of the aforementioned, outdated methods, such as the counting of postulates and the superficial estimation of the simplicity of formulae. 
This means that the new calculation method also has to capture all of the information contained within all of the postulates, axioms, and definitions, that were required to give a precise meaning to the mathematical formulae of the physicist's model.
While unaware of the stagnation's aforedescribed root cause, Chaitin, 
who worked on metamathematics (math about math), 
noted that the correct definition of the complexity of a formal axiomatic system should be "taken to be the minimal size of a self-delimiting program for enumerating the set of theorems of the formal system"
\cite{chaitin1992information}.
Of course, the axioms have to be encoded within this program, 
rather than being provided to the program via some input.

So, Chaitin's method will yield the amount of information in the axiomatic system, but what about information in the postulates and definitions?
This is easy:
One can treat the formal definitions and postulates as if they were axioms and one can add them to the set of axioms.
Then, Chaitin's method will obviously additionally also capture the amount of information in the definitions and postulates.
Since the complete the model has to return predicted observations, 
the model would be completed by adding an algorithm that is purposed to leverage some of the computed theorems to compute and output such predicted observations.
The total model complexity would then be size of this algorithm plus the size of the algorithm described by Chaitin.
One such a calculation could then, in principle, hypothetically, 
be leveraged to either support or debunk string theory,
as Hutter noted \cite{hutter2013subjective}.
This would be achieved by calculating how much more complex (or simpler) string theory is, as compared to the standard model plus general relativity.
Hypothetically, string theory then might be debunked due the potentially excessive complexity of its sophisticatedly curled-up, high-dimensional manifolds. 

Be that as it may, some physicists now may rightly be concerned about the unnaturalness of programming languages,
since Hutter mentions the programming language C and Chaitin uses the language LISP to measure complexity. 
But worry not, as we have already completed the discussion of this issue earlier on, in \mbox{Subsection \ref{reference formalism section}},
where we concluded that the optimal choice of the language (the reference formalism) would be the good old-fashioned and widely-known, untyped lambda calculus (ULC), 
which is orders of magnitude simpler\footnote{
This is not based on circular reasoning,
see \mbox{Subsubsections \ref{circular} and \ref{circular resolution}}.
}
than C or LISP,
and thus, the ULC may also be considered to be much less contrived, or much more natural if you will.

% Paragraph explaining the feasibility and metamath, based on look up tables.
As previously mentioned,
one needs to capture all of the information contained within all of the postulates, axioms, and definitions, that were required to give a precise meaning to the mathematical formulae of a physicist’s model.
This will involve the foundations of mathematics \cite{russell2020principles,kleene1952introduction},
such as the inference rules of first-order logic, the axioms of set theory, the axioms/definition of real numbers, and so forth---and all of this will have to be translated into the ULC (or BLC), 
which appears to be a dauntingly large and difficult task.
If one were to demand each physicist to use such a method to calculate their own model complexities from scratch on their own, this would obviously be deemed to be an infeasibly time-consumptive method.
However, since most of their models are based on the same foundations of mathematics, 
metamathematicians should first quantify the information required for the many common mathematical definitions that physics usually relies on. 
The metamathematicians would then release the resulting numbers in the form of lookup tables that would enable each physicist to assemble their model complexity calculations efficiently, 
since each physicist will then only need to sum up relatively the few complexities corresponding to the few different components of their physics formulae.

Here is a short example. 
A physicist's new model may consist of multiple integrals.
The physicist would then access the lookup table to look up the complexity of a general definition of integrals. 
The physicist would then incorporate this complexity into the calculation of the model's total complexity.
The physicist would go on to add the small complexities required to reference this definition multiple times, 
rather than repeatedly adding of the complexity of the definition.
Any definition of the integral is built on the definition of $\lim$.
Therefore, if the physics formula furthermore contains a derivative, 
the physicist would look up and incorporate its complexity but minus the complexity of the definition of $\lim$, as it had already been accounted for as part of the integral's definition.
If the physics formula contains a numeric constant, its complexity would be calculated based on the precision of the constant, 
e.g., the minimal size of the numeric constant when represented in a binary number system.
And so on and so forth.
Even decades in the past, equipped with only pen and paper, 
such a methodology based on lookup tables, 
would already have been manageable enough for it to be incorporated it into all fundamental theoretical physics modeling.
So then why had this methodology not been practiced decades in the past already?
The reason is simply that, before our paper,
as valuable as the various previous authors' contributions may have been,
none of them had resolved all of the objections resolved in our paper nor had they made all of the necessary conceptual connections.

Meanwhile, in our modern day and age, 
it has become be even easier to adopt such a methodology since,
conveniently and coincidentally, the foundations of mathematics have already been translated into the ‘Lean’ language \cite{ebner2017metaprogramming,lean2020library,de2015lean}, 
which is the most widely used functional programming language intended for mathematical proof checking.
Furthermore, this continuously growing, digital library \cite{lean2020library,maintaining2020library}, 
presently, already contains many definitions
pertaining to differential/integral calculus, complex analysis, linear algebra, and smooth manifolds. 
This means that meanwhile, a great fraction of the relevant metamathematical work has already been completed and digitized.
So, rather than printing booklets with lookup tables, 
it would be smarter to develop novel software for this purpose.
The physicists would then simply type their formulae into the software that will automatically calculate model the complexity for them.
Moreover, the metamathematical work of composing the ‘complexity lookup tables’ should also be automated away by automating the conversion from Lean formulae into a ULC/BLC formula/program, 
in concordance with Chaitin's aforementioned method \cite{chaitin1992information},
which would then enable the efficient, comprehensive, and objective quantification of all the algorithmic complexity hidden within sets of formulae that constitute a typical theoretical physicist's model.
From our explanation, it should have become apparent, by now, that this methodology would be technically and practically feasible.

The following short paragraph concerns a minor technical detail:
To define the size of ULC formulae in bits or nats, there exist multiple options.
One readily available option is to rely on Tromp's binary lambda calculus (BLC) \cite{tromp2007binary,tromp2023functional},
where the ULC is converted into binary code,
so that the number of bits can then simply be counted directly. 
Another option, would be to endow the ULC with a simple stochastic grammar that assigns a probability to any ULC formula,
such that the binary/natural logarithm of that probability would yield the size in bits/nats.
This would avoid the forced binarization of the ULC, 
which some may consider unnatural.
There would be minor differences between the results of these options, 
such that ultimately, these details would not matter much.
Further details of this sort are outside the scope of this paper.

%Perfect Recap of the stagnation:
Recall that the primary root cause of the stagnation was the physicists' present reliance on poor methodology for estimating model complexities.
As explained in the previous subsection,
due to the overwhelmingly large number of theoretical physics papers, 
the relatively few papers containing significant novel theoretical insights, 
are essentially buried underneath a heaps of insignificant papers,
and without a ranking based on model complexity,
the identification of significant papers is analogous to finding needles in a haystack, 
where one first needs to spend considerable amounts of time to study each hay straw individually, i.e., 
it is infeasibly inefficient to find the theoretically significant papers.
This then leads to overall poorly informed choices on the allocation of various resources, 
on multiple economic levels of analysis.

Since the new methodology would repair this defective state of affairs, 
any new fundamental theoretical physics papers should be required to report their model complexities, 
along with the exact research problems which their models are addressing.
In order to further increase the transparency and efficiency of the model complexity-based rankings of papers,
all of the different research problems should become standardized across the field of fundamental theoretical physics,
such that for each of these standard problems separately, 
its solutions could be ranked officially.
A list of standardized problems could possibly be maintained by a small, neutral, international organization of physicists, such as IUPAP.
Due to the current lack of complete ToEs, 
it will be insufficient to merely standardize the ultimate problem of finding a ToE 
and to then hope that the Hutter's ‘complete ToE selection principle’ \cite{hutter2010complete_occamProof} would start to take its effect.
Instead, as explained in \mbox{Subsection \ref{hierarchy}}, 
such problems form a hierarchy that is to be tackled effectively via a hierarchical strategy.
Therefore, the official standardization of research problems must start from the smallest subproblems, 
such as the modeling of the simplest symmetry groups, 
and continue through multiple intermediate levels, 
all the way up to the problem of finding the correct ToE. 
But of course, if the many subproblems would not become standardized, 
it would complicate the comparisons between papers, leading to a suboptimal rate of progress.
Speaking of standards, 
it may also be useful to publish exemplary calculations of the present-day complexities of the standard model and of general relativity, 
that would serve as a standard reference for calculating complexities of similar alternative models.

% Regularization, the term
For the physics community, it will be noteworthy and interesting that the scholarly lineage of this new metamathematical methodology, 
if traced 70 years back into the past,
turns out to have roots in the field of physics itself, 
namely, regularization \cite{pauli1949invariant,levi1915sulla}.
The term ‘regularization’ is widely-known and widely-used in the fields of statistics and machine learning to refer to the penalization of excessive model complexity \cite{tian2022comprehensive}
and was introduced by Tikhonov \cite{tikhonov1963solution,tikhonov1965use}, 
who imported this term from physics, before its stagnation.
Since the aforedescribed new methodology would regularize the physics research field metascientifically, 
and since this regularization involves metamathematics,
we will henceforth refer to this new methodology as ‘metamathematical regularization'.
A similar but more restricted approach was once also employed by Soklakov \cite{soklakov2002occam} to retrodict aspects of known laws of physics.
As we will touch on next, the full extent of the implications of metamathematical regularization could be much deeper than the mere retrodiction of known laws as well as much deeper than the mere refutation of string theory.

%Plausibility of discrete models
According to Lev \cite{lev2020finite,lev2020my}, 
the stagnation the research field is actually due to quantum theory's reliance on 
"classical mathematics involving the notions of infinitely small/large and continuity".
This means that, at the most fundamental level, 
physics may be fully discrete and should be modeled accordingly to alleviate the stagnation,
while anything continuous would have to be emergent.
Could metamathematical regularization corroborate such a claim?
In principle, yes, since continuous models are built on top of the foundations of mathematics that include first-order logic, 
the axioms of set theory, the axioms/definition of real numbers, and so forth,
and such a sophisticated foundation would not be necessary for fully discrete models. 
To add some more detail, continuous models rely on the formal axiomatic system at the traditional foundations of mathematics, which is a particular kind of ‘rewriting system',
where each rewriting step results in a new logical statement or theorem.
However, for fully discrete models, different, better suited rewriting systems can be employed, where each rewriting step would correspond directly to an event in spacetime.
This means that the regularization of fully discrete models does not require the Lean library \cite{lean2020library} and does not even require predicate logic at the most fundamental level. 
To determine the model complexity for regularization purposes, 
one would just translate the rewriting system directly into the ULC, which is itself also already a rewriting system.
This means that fully discrete models do posses at least the potential to be much simpler than continuous models.
Not only does this potential simplicity imply that discrete models have a relatively high chance of leading to success, but furthermore, 
the adoption of the new regularization procedure will be much easier because Lean will not be necessary, as just explained. 
However, due to the absence of the traditional tool set of continuous math at the most fundamental level, 
such fully discrete modeling will be more challenging, but could, in turn, 
yield more impressive solutions, if regularized correctly. 
Various famous, fully discrete \cite{t2016cellular,wolfram2020project} or partially discrete \cite{rovelli2008loop,surya2019causal} approaches exist, 
but all of them would still need to start regularizing themselves metamathematically first, before unlocking their real potential.
For all these reasons, such discrete approaches should be investigated closely by physicists looking to get a head start at present. 

%Lambda Calculus Naturality for discrete spacetime
Be that as it may, even if one were to adhere to continuous models only, 
metamathematical regularization is also the key to achieving progress with just continuous models.
But as a final sidenote on discrete models of spacetime,
some physicists may be concerned about the fact that spacetime events are only partially ordered in time, while in their experience, computational steps are usually totally ordered in time.
However, the computational/rewriting steps in ULC are also partially ordered rather than totally ordered, similar to events in spacetime.
Yet, since ULC is usually represented algebraically, as a single sequence of symbols, 
physicists may furthermore be concerned that this apparent ‘one-dimensionality’ of the sequence may render ULC unfit to represent the microstructure of 3D space.
This detail is outside the scope of our paper, 
but suffice to say that this is not a real problem since a bound variabled can be located arbitrarily many symbols away from the respective heads that they are bound to,
thus naturally enabling the representation of arbitrary networks of relations, 
including the likes of the microstructure of 3D space.
This concludes our discussion of discrete models.

%Turns out, physics is not really overfitting
It is generally known that the neglect of regularization procedures permits overfitting \cite{hawkins2004problem,nannen2010short},
which will tend to lead to false predictions in novel domains.
This is not only true for curve fitting but also for scientific modeling in general, 
as should have become apparent from our previous sections.
One therefore may assume that the field of fundamental theoretical physics is stagnant due to overfitting.
However, this is not quite true, since overfitting would imply that better, simpler models would not fit the old experimental data quite as closely as the incorrect, complicated models. 
Nevertheless, when faulty experimental data was released, 
this immediately lead to numerous new theoretical models, resulting in numerous false predictions \cite{hossenfelder2018inside,hossenfelder2018lost},
which could actually be interpreted as overfitting,
since in this case, the correct models would indeed not fit the data quite as closely as the false models.
But speaking of false predictions, even without faulty data and without overfitting, 
the lack of metamathematical regularization has still led to various false predictions \cite{rovelli2019dangers}.

%Paragraph on inclusion of Empirical Loss
One more detail to be mentioned is that regularization is usually understood to involve the addition of a regularization term (model complexity) to the empirical loss (cross entropy), 
resulting in the regularized loss,
as we have explained at the end of \mbox{Subsection \ref{stochastic section}}.
So, when addressing a standardized problem, 
the fundamental theoretical physics papers should report the empirical loss alongside the model complexity.
The details of how to calculate the empirical loss will depend on the problem being addressed, 
but such calculations should be easier than to calculate the model complexity, 
so, we will not address this topic further.
Let us instead move toward concluding this section.

%Motivational paragraph
Who in the field of physics should be most motivated to adopt metamathematical regularization?
We will answer this question while climbing through the levels of analysis again \cite{dopfer2004micro}.
Firstly, any physicist who believes their own models to be closer to reality than alternative models should be motivated to adopt the new methodology,
since it would allow them to calculate and publish the low complexities of their own models in order to lay claim to having found the correct solution to an open problem---at least as long as no other physicist rises to the challenge of finding an even simpler model. 
Next, research communities that have long been working within a promising theoretical framework should be motivated to adopt the new methodology,
because it would allow for the identification of the strengths and weaknesses of their theories, 
conducing the import of solutions from other communities, to plug the holes of their theories,
thereby strengthening their own theoretical framework in the long run. 
Also, scientific journals should be motivated to require the new methodology for publication, 
as it would allow them to more easily distinguish  significant papers from the many insignificant papers, when deciding on which papers to publish,
thereby increasing the journal's impact factor in the long run.
Lastly, the same will be true for large research-funding agencies, 
as the transparency brought about by the new methodology would allow them to more accurately compare the potential merit of different research directions, 
enabling a smarter allocation of resources. 

%Final summary of the whole section
To summarize what we have explained throughout these three subsections: 
In fundamental theoretical physics research,
the adoption of metamathematical regularization as a methodological quality standard, 
in combination with the official standardization of theoretical research subproblems, 
could increase the chances of this field's forthcoming success by orders of magnitude. 
Moreover, we have shown the adoption of these new methodologies to be practically and technically feasible at present.

\section{Conclusion}\label{conclusion}
Throughout this paper, 
we have journeyed through discussions on many topics,
ranging from the simplest formalisms all the way up to the complicated ecosystem consisting of all theoretical physics researchers.  
Thereby, we have resolved all twelve explicitly stated objections, 
plus many implicit objections spread throughout our text.
Concerning Occam's razor, we have made sure to eliminate any reasonable doubt that we have come across. 
As we have shown, a modernized, highly general Occam's razor can be proven mathematically, 
starting from only reasonable initial assumptions (see \mbox{Subsection \ref{assumptions}}). 
One key takeaway is that the power of Occam's razor as a guiding principle stems from the fact that relatively small differences in complexity can already suffice for the ratios between probabilities to become hugely lopsided, due to certain exponential relations (See \mbox{Equations \ref{odds exp equation} and \ref{odds computable}}). 
We have then explained at length how this great power could be harnessed by the fundamental theoretical physics research community
via the practically feasible adoption of a new ‘upgrade’ to their methodology, namely, ‘metamathematical regularization',
in combination with the standardization of their research subproblems. 
This upgrade would easily increase the physicists' chances of success by orders of magnitude.
The next step would be for experts from the fields of algorithmic information theory, metamathematics, metascience, and theoretical physics to come together to devise a plan for the embarkation into an age of discovery via metamathematical regularization, 
such that the old stagnation may hopefully be alleviated in the near future.

\subsection*{Acknowledgments and Declarations}
I thank the Sejny Summer Institute 2023 for allowing me to engage in fruitful discussions with theoretical physicists for one week, 
in person, all expenses covered.
These valuable interactions allowed me to diagnose the root cause of their field's stagnation.
Subsequently, I realized that the old mathematical arguments for Occam's razor needed to be improved, which led me to write the present paper in the subsequent years.
\\ Declarations:
No artificial intelligence was used to generate any part of this manuscript.
\\ This research has not received any funding. 
There is no conflict of competing interests.

\bibliographystyle{unsrt}
%\bibliography{references}

% Bibliography adapted to arXiv's requirements:

\end{document}